\documentclass[12pt]{article}
\usepackage{cite}
\usepackage{graphicx,feynarts,amsmath,amssymb}

\makeatletter
\def\citer{\@ifnextchar [{\@tempswatrue\@citexr}{\@tempswafalse\@citexr[]}}
 
\def\@citexr[#1]#2{\if@filesw\immediate\write\@auxout{\string\citation{#2}}\fi
  \def\@citea{}\@cite{\@for\@citeb:=#2\do
    {\@citea\def\@citea{--\penalty\@m}\@ifundefined
       {b@\@citeb}{{\bf ?}\@warning
       {Citation `\@citeb' on page \thepage \space undefined}}%
\hbox{\csname b@\@citeb\endcsname}}}{#1}}
\makeatother

\def\refeq#1{\mbox{Eq.~(\ref{#1})}}
\def\refeqs#1{\mbox{Eqs.~(\ref{#1})}}
\def\reffi#1{\mbox{Fig.~\ref{#1}}}
\def\reffis#1{\mbox{Figs.~\ref{#1}}}

\def\refse#1{\mbox{Sect.~\ref{#1}}}

\def\citere#1{\mbox{Ref.~\cite{#1}}}
\def\citeres#1{\mbox{Refs.~\cite{#1}}}

\newcommand{\ie}{i.e.\ }
\newcommand{\FA}{\textit{FeynArts}}
\newcommand{\FC}{\textit{FormCalc}}
\newcommand{\LT}{\textit{LoopTools}}
\newcommand{\ri}{\mathrm{i}}

\def\xtilde#1{%
  \setbox0\hbox{$\tilde#1$}%
  \rlap{\raise\ht0\hbox{\tiny$_{\,(\;\,)}$}}%
  \tilde#1%
}

\newcommand{\At}{A_t}
\newcommand{\Ab}{A_b}
\newcommand{\Atau}{A_\tau}
\newcommand{\Xt}{X_t}

\newcommand{\msusy}{M_{\text{SUSY}}}

\newcommand{\Pe}{\phi_1}
\newcommand{\Pz}{\phi_2}

\newcommand{\PePz}{\phi_1\phi_2}
\newcommand{\mpe}{m_{\Pe}}
\newcommand{\mpz}{m_{\Pz}}
\newcommand{\mpez}{m_{\PePz}}

\newcommand{\msbar}{$\overline{\rm{MS}}$}
\newcommand{\msbarm}{\overline{\rm{MS}}}

\def\order#1{${\cal O}(#1)$}
\newcommand{\cp}{{\cal CP}}

\newcommand{\cM}{{\cal M}}

\newcommand{\wz}{\sqrt{2}}
\newcommand{\edz}{\frac{1}{2}}
\def\ed#1{\frac{1}{#1}}
\newcommand{\twol}{two-loop}
\newcommand{\onel}{one-loop}

\newcommand{\fh}{{\em FeynHiggs}}

\newcommand{\MW}{M_W}
\newcommand{\MZ}{M_Z}
\newcommand{\MA}{M_A}
\newcommand{\mh}{m_h}
\newcommand{\mH}{m_H}
\newcommand{\Mh}{M_h}
\newcommand{\MH}{M_H}
\newcommand{\HSM}{H_{\rm SM}}
\newcommand{\MHSM}{M_{\HSM}}

\newcommand{\mhH}{m_{hH}}

\newcommand{\MHp}{M_{H^\pm}}
\newcommand{\mhmax}{m_h^{\rm max}}
\newcommand{\siHenh}{\si_H^{\text{enh}}}

\newcommand{\mt}{m_{t}}

\newcommand{\mb}{m_{b}}

\newcommand{\mgl}{m_{\tilde{g}}}

\newcommand{\Stop}{\tilde{t}}
\newcommand{\StopL}{\tilde{t}_L}
\newcommand{\StopR}{\tilde{t}_R}
\newcommand{\Stope}{\tilde{t}_1}
\newcommand{\Stopz}{\tilde{t}_2}

\newcommand{\Sbot}{\tilde{b}}

\newcommand{\Sbote}{\tilde{b}_1}
\newcommand{\Sbotz}{\tilde{b}_2}

\newcommand{\Stau}{\tilde{\tau}}
\newcommand{\Staue}{\tilde{\tau}_1}
\newcommand{\Stauz}{\tilde{\tau}_2}
\newcommand{\Sneut}{\tilde{\nu}_\tau}

\newcommand{\tsf}{\theta\kern-.20em_{\tilde{f}}}
\newcommand{\tsfp}{\theta\kern-.20em_{\tilde{f}\prime}}
\newcommand{\tsq}{\theta\kern-.15em_{\tilde{q}}}
\newcommand{\sw}{s_W}
\newcommand{\cw}{c_W}

\newcommand{\KL}{\left(}
\newcommand{\KR}{\right)}
\newcommand{\KKL}{\left[}
\newcommand{\KKR}{\right]}
\newcommand{\KKKL}{\left\{}
\newcommand{\KKKR}{\right\}}

\newcommand{\VL}{\left( \begin{array}{c}}
\newcommand{\VR}{\end{array} \right)}
\newcommand{\ML}{\left( \begin{array}{cc}}
\newcommand{\MLd}{\left( \begin{array}{ccc}}
\newcommand{\MLv}{\left( \begin{array}{cccc}}
\newcommand{\MR}{\end{array} \right)}

\newcommand{\re}{\mathop{\rm Re}}

\newcommand{\tb}{\tan \beta}

\newcommand{\sinb}{\sin \beta\hspace{1mm}}

\newcommand{\SQb}{\sin^2\beta\hspace{1mm}}

\newcommand{\Cb}{\cos \beta\hspace{1mm}}
\newcommand{\CQb}{\cos^2\beta\hspace{1mm}}

\newcommand{\Sa}{\sin \alpha\hspace{1mm}}

\newcommand{\Ca}{\cos \alpha\hspace{1mm}}

\newcommand{\Sba}{\sin (\beta - \alpha)}
\newcommand{\Cba}{\cos (\beta - \alpha)}

\newcommand{\CZa}{\cos 2\alpha\hspace{1mm}}

\newcommand{\CZb}{\cos 2\beta\hspace{1mm}}

\newcommand{\tev}{\,\, \mathrm{TeV}}
\newcommand{\gev}{\,\, \mathrm{GeV}}
\newcommand{\mev}{\,\, \mathrm{MeV}}

\newcommand{\BC}{\begin{center}}
\newcommand{\EC}{\end{center}}
\newcommand{\BE}{\begin{equation}}
\newcommand{\EE}{\end{equation}}
\newcommand{\BEA}{\begin{eqnarray}}
\newcommand{\BEAnn}{\begin{eqnarray*}}
\newcommand{\EEA}{\end{eqnarray}}
\newcommand{\EEAnn}{\end{eqnarray*}}
\newcommand{\non}{\nonumber}
\newcommand{\id}{{\rm 1\kern-.12em
\rule{0.3pt}{1.5ex}\raisebox{0.0ex}{\rule{0.1em}{0.3pt}}}}

\newcommand{\gf}{G_F}

\def\al{\alpha}
\def\aeff{\al_{\rm eff}}

\def\be{\beta}

\def\ga{\gamma}

\def\de{\delta}

\def\si{\sigma}

\def\Ga{\Gamma}

\def\Ghn{\Ga_h^{(0)}}
\def\GHn{\Ga_H^{(0)}}

\def\De{\Delta}

\def\Si{\Sigma}

\def\hSi{\hat{\Sigma}}
\def\hSip{\hat{\Sigma}'}

\def\hSiH{\hSi_{HH}}
\def\hSih{\hSi_{hh}}
\def\hSihH{\hSi_{hH}}
\def\hSipH{\hSip_{HH}}
\def\hSiph{\hSip_{hh}}

\newcommand{\eennh}{$e^+e^- \to \bar\nu \nu \, h$}
\newcommand{\eennH}{$e^+e^- \to \bar\nu \nu \, H$}
\newcommand{\eeneneH}{$e^+e^- \to \bar\nu_e \nu_e \, H$}
\newcommand{\eennhH}{$e^+e^- \to \bar\nu \nu \, \{h,H\}$}
\newcommand{\eenenehH}{$e^+e^- \to \bar\nu_e \nu_e \, \{h,H\}$}

\newcommand{\fb}{\mbox{~fb}}

\newcommand{\iabm}{\mbox{ab}^{-1}}

\newcommand{\hbb}{h \to b\bar{b}}

\newcommand{\htautau}{h \to \tau^+\tau^-}

\newcommand{\WWH}{WW \to H}

\newcommand{\rZh}{\hat Z_h}

\newcommand{\wrZh}{\sqrt{\rZh}}

\newcommand{\rZH}{\hat Z_H}

\newcommand{\wrZH}{\sqrt{\rZH}}

\newcommand{\rZhH}{\hat Z_{hH}}

\newcommand{\rZHh}{\hat Z_{Hh}}

\graphicspath{{plots/}}

\allowdisplaybreaks

\begin{document}
\thispagestyle{empty}

\def\thefootnote{\fnsymbol{footnote}}

\begin{flushright}
DCPT/02/120\\ 
IPPP/02/60\\
LMU 08/02\\
MPI-PhT/2002-33\\
hep-ph/0211204
\end{flushright}

\vspace{1cm}

\begin{center}

{\large\bf MSSM Higgs-Boson Production at the Linear Collider:}

\vspace{0.4cm}

{\large\bf\boldmath Dominant Corrections to the $WW$-Fusion Channel}
 
\vspace{1cm}

{\sc 
T.~Hahn$^{1}$%
\footnote{email: hahn@feynarts.de}%
, S.~Heinemeyer$^{2}$%
\footnote{email: Sven.Heinemeyer@physik.uni-muenchen.de}%
, and G.~Weiglein$^{3}$%
\footnote{email: Georg.Weiglein@durham.ac.uk}
}

\vspace*{1cm}

{\sl
$^1$Max-Planck-Institut f\"ur Physik (Werner-Heisenberg-Institut),
F\"ohringer Ring 6, \\
D--80805 Munich, Germany

\vspace*{0.4cm}

$^2$Institut f\"ur Theoretische Elementarteilchenphysik,
LMU M\"unchen, Theresienstr.\ 37, D--80333 Munich, Germany

\vspace*{0.4cm}

$^3$Institute for Particle Physics Phenomenology, University of Durham,\\
Durham DH1~3LE, UK

}

\end{center}

\vspace*{1cm}

\begin{abstract}
In the Minimal Supersymmetric Standard Model (MSSM) we calculate the
corrections to neutral $\cp$-even Higgs-boson production in the
$WW$-fusion and Higgs-strahlung channel, \eennhH, at a future Linear Collider,
taking into account all
\order{\al} corrections arising from loops of fermions and sfermions.  
For the production of the lightest MSSM Higgs boson, $h$, we find genuine
loop corrections (beyond the universal Higgs propagator corrections) of
up to $-5\%$.  For the heavy $\cp$-even neutral Higgs boson, $H$, which
shows decoupling behavior at tree level, we find non-negligible
corrections that can enhance the cross section considerably in parts of
the MSSM parameter space. At a center-of-mass energy of $\sqrt{s} = 1000
\gev$, heavy $\cp$-even Higgs-boson masses of up to $\MH \lesssim 700
\gev$ are accessible at the Linear Collider in favorable regions of the 
MSSM parameter space.
\end{abstract}

\def\thefootnote{\arabic{footnote}}
\setcounter{page}{0}
\setcounter{footnote}{0}

\newpage


\section{Introduction}

Disentangling the origin of electroweak symmetry breaking is one of the main tasks
of the current and next generation of colliders. The prime candidate is a Higgs
mechanism with elementary scalar particles below the TeV scale. Within the
electroweak Standard Model (SM) the minimal version of the Higgs mechanism is
implemented, i.e.\ one doublet of complex scalar fields giving rise to one physical
Higgs boson. On the other hand, theories based on Supersymmetry (SUSY) \cite{susy}
are widely considered as the theoretically most appealing extension of the SM.
Contrary to the SM, two Higgs doublets are required in the minimal realization,
resulting in five physical Higgs bosons~\cite{hhg}. The Higgs sector of the Minimal
Supersymmetric Standard Model (MSSM) can be expressed at lowest order in terms of
$\MZ$, $\MA$ (the mass of the $\cp$-odd Higgs boson), and $\tb = v_2/v_1$, the ratio
of the two vacuum expectation values. While the discovery of one light Higgs boson
might well be compatible with the predictions both of the SM and the MSSM, the
discovery of one or more other heavy Higgs boson would be a clear and unambiguous
signal for physics beyond the SM.

In the decoupling limit, \ie for $\MA\gtrsim 200 \gev$, the heavy MSSM
Higgs bosons are nearly degenerate in mass, $\MA \sim \MH \sim \MHp$.  The
couplings of the neutral Higgs bosons to SM gauge bosons are
proportional to
\begin{equation}
\begin{gathered}
VVh \sim VHA \sim \Sba\,, \\
VVH \sim VhA \sim \Cba\,,
\end{gathered}
\qquad (V = Z, W^\pm)
\end{equation}
where $\al$ is the angle that 
diagonalizes the $\cp$-even Higgs sector.
In the decoupling limit one finds $\be - \al \to
\pi/2$, \ie $\Sba \to 1$, $\Cba \to 0$.

At the LC, the possible channels for neutral Higgs-boson production are
the production via $Z$-boson exchange,
\begin{equation}
\begin{aligned}
e^+e^- &\to Z^* \to Z \{h,H\}\,, \\
e^+e^- &\to Z^* \to A \{h,H\}\,,
\end{aligned}
\end{equation}
and the $WW$-fusion channel,
\begin{equation}
e^+e^- \to \bar\nu_e W^+ \; \nu_e W^- \to \bar\nu_e \nu_e \{h,H\}\,.
\label{eq:eennhH}
\end{equation}
As a consequence of the coupling structure, in the decoupling limit the heavy Higgs
boson can only be produced in $(H,A)$~pairs. This limits the LC reach to
$\MH\lesssim\sqrt s/2$. Higher-order corrections to the $\WWH$ channel from loops of
fermions and sfermions, however, involve potentially large contributions from the
top and bottom Yukawa couplings and could thus significantly affect the decoupling
behavior.

In this paper we have evaluated the one-loop corrections of fermions and sfermions
to the process \eennhH, i.e.\ to the production of a neutral $\cp$-even Higgs boson
in association with a neutrino pair both via the $WW$-fusion and the Higgs-strahlung
mechanism. In the latter case the $Z$~boson is connected to a neutrino pair, $e^+e^-
\to Z \{h,H\} \to \bar\nu_l\nu_l \{h,H\}$, with $l = e, \mu, \tau$ (where the latter
two neutrinos result in an indistinguishable final state in the detector). Our
results have been derived using the packages {\em FeynArts}, {\em FormCalc}, and
{\em LoopTools}~\cite{feynarts,formcalc}.

While the well-known universal Higgs-boson propagator corrections turned out not to
significantly modify the decoupling behavior of the heavy $\cp$-even Higgs boson, an
analysis of the process-specific contributions to the $WWH$ vertex has been missing
so far. Taking into account all loop and counter-term contributions to the process
\eennhH\ from fermions and sfermions and including also the effects of beam
polarization in our analysis, we investigate in this paper the LC reach for the
heavy $\cp$-even Higgs boson. We have obtained results for values of the MSSM
parameters according to the four benchmark scenarios defined in
\citere{LHbenchmark}. While within these benchmark scenarios we find that the loop
corrections do not significantly enhance the LC reach for heavy $\cp$-even Higgs
boson production and in some cases even slightly reduce the accessible parameter
space, we have also investigated MSSM parameter regions where the loop effects do in
fact lead to a significant improvement of the LC reach. In ``favorable'' MSSM
parameter regions an $e^+e^-$ LC running at $\sqrt{s} = 1$~TeV can be capable of
producing a heavy $\cp$-even Higgs boson with a mass up to $\MH \lesssim 700$~GeV.

Concerning the production of the light $\cp$-even Higgs boson, an
accurate prediction of the production cross section for precision
analyses will be necessary. Aiming for analyses at the percent
level~\cite{eennHexp} also requires a prediction of the production cross
section in this range of precision. Besides the already known universal
Higgs propagator corrections, in particular loops from fermions and
sfermions (especially from the third family) are expected to give
relevant contributions. We analyze our results for the parameters of the
four benchmark scenarios defined in \citere{LHbenchmark} and study the
results as a function of different SUSY parameters. We discuss the relative
importance of the fermion- and the sfermion-loop contributions and
furthermore evaluate the fermion-loop correction within the SM for
comparison purposes.

Electroweak loop effects on processes within the MSSM where a single Higgs 
boson is produced have recently drawn considerable interest in the
literature, and we compare our results to existing ones where there is
overlap. 
Within the SM, the tree-level results for Higgs-boson production in the
$WW$-fusion channel are available for several years
already~\cite{eennHtree0,eennHtree}. On the other hand, evaluating the
full one-loop corrections within the SM has been attempted only very
recently~\cite{WWHSM_RC2002}. In the MSSM, corrections from
third-generation fermions and sfermions to \eennh\ have been presented
in \citere{Wiener}. Production of the heavy $\cp$-even Higgs boson has
only been considered in \citere{Wiener} for small values of $\MA$, where 
the heavy $\cp$-even Higgs boson has SM-like couplings, while the
decoupling region has not been investigated. We have compared our
results for the production of the light $\cp$-even Higgs boson
(restricted to the corrections from the third generation only) with the
ones given in \citere{Wiener} and find significant deviations.
While the authors of \citere{Wiener} find large corrections from the
loops of third generation fermions both in the SM and in the MSSM, 
the correction from this class of diagrams in our result turns out to be
much smaller and does not exceed $\pm 2$\%.

Electroweak loop effects have also been evaluated for other processes
with single Higgs-boson production within the MSSM. The process $e^+e^-
\to \nu_e \bar \nu_e A$ has been evaluated at the full \onel\ level in
\citere{eennA}, where the cross section has been found to be too small
for $A$ detection at the LC via this channel. The results for the
Higgs-strahlung process, $e^+e^- \to Z^* \to Z\{h,H\}$, and the
associated production, $e^+e^- \to Z^* \to A\{h,H\}$, containing the
complete one-loop contributions and the leading \twol\ corrections
entering via Higgs-boson propagators, have been given in \citere{eehZhA}.  
The \twol\ corrections have been found to yield corrections at the
5--10\% level. As explained above, these processes are not suitable for
heavy Higgs boson production with $\MA > \sqrt{s}/2$. Furthermore, the
production of a charged Higgs boson in association with a $W$~boson has
been evaluated, including the full \onel\
corrections~\cite{logan&su,WH_other}. It has been found that this channel
possesses only a small potential for charged Higgs boson production with
$\MHp > \sqrt{s}/2$.

The rest of the paper is organized as follows: In
\refse{sec:MSSMhiggs} we review the necessary features of the MSSM
Higgs sector. Details about the calculation are presented in
\refse{sec:WWhH}. The numerical analysis for light and heavy
$\cp$-even Higgs boson production, the corresponding SM
result, and the comparison with existing results can be found in
\refse{sec:numeval}. We conclude with \refse{sec:conclusinos}.


\section{The MSSM Higgs sector}
\label{sec:MSSMhiggs}

At the tree level the mass matrix of the neutral $\cp$-even Higgs bosons
in the $\phi_1,\phi_2$ basis can be expressed in terms of $\MZ$, $\MA$
(the mass of the $\cp$-odd Higgs boson), and $\tb = v_2/v_1$, the ratio
of the two vacuum expectation values, as follows~\cite{hhg}:
\begin{align}
M_{\rm Higgs}^{2, {\rm tree}}
&= \begin{pmatrix}
   \mpe^2 & \mpez^2 \\ 
   \mpez^2 & \mpz^2
   \end{pmatrix} \non\\
&= \begin{pmatrix}
   \MA^2 \SQb + \MZ^2 \CQb & -(\MA^2 + \MZ^2) \sinb \Cb \\
   -(\MA^2 + \MZ^2) \sinb \Cb & \MA^2 \CQb + \MZ^2 \SQb
   \end{pmatrix}.
\end{align}
Transforming to the mass-eigenstate basis yields 
\begin{equation}
M_{\rm Higgs}^{2, {\rm tree}} 
  \overset{\al}{\longrightarrow}
  \ML \mH^2 & 0 \\ 0 &  \mh^2 \MR,
\end{equation}
$\mh$ and $\mH$ being the tree-level masses of the neutral
$\cp$-even Higgs bosons and 
\begin{align}
\begin{pmatrix} H \\ h \end{pmatrix}
&= \begin{pmatrix}
   \Ca & \Sa \\
   -\Sa & \Ca
   \end{pmatrix}
   \begin{pmatrix} \phi_1 \\ \phi_2 \end{pmatrix}.
\label{eq:higgsrotation}
\end{align}
The mixing angle $\alpha$ is related to $\tb$ and $\MA$ by
\begin{equation}
\tan 2\alpha = \tan 2\beta\,\frac{\MA^2 + \MZ^2}{\MA^2 - \MZ^2}\,,
\quad -\frac{\pi}{2} < \alpha < 0\,.
\label{alphatree}
\end{equation}

\bigskip

In the Feynman-diagrammatic approach, the higher-order-corrected masses
of the two $\cp$-even Higgs bosons, $\Mh$ and $\MH$, are derived beyond
tree level by determining the poles of the $h$--$H$-propagator matrix
whose inverse is given by
\begin{equation}
\left(\Delta_{\rm Higgs}\right)^{-1}
= -\ri\begin{pmatrix}
  q^2 - \mH^2 + \hSi_{H}(q^2) & \hSi_{hH}(q^2) \\
  \hSi_{hH}(q^2) & q^2 - \mh^2 + \hSi_{h}(q^2)
  \end{pmatrix},
\label{eq:higgspropagatormatrixnondiag}
\end{equation}
where the $\hSi$ denote the renormalized Higgs-boson self-energies.
Determining the poles of the matrix $\Delta_{\rm Higgs}$ in
\refeq{eq:higgspropagatormatrixnondiag} is equivalent to solving
the equation
\begin{equation}
\left[q^2 - \mh^2 + \hSi_{hh}(q^2) \right]
\left[q^2 - \mH^2 + \hSi_{HH}(q^2) \right] - 
\left[\hSi_{hH}(q^2)\right]^2 = 0\,.
\end{equation}
The renormalized Higgs-boson self-energies are given by
\begin{equation}
\begin{aligned}
\hSi_{HH}(q^2)
&= \Si_{HH}(q^2) + \de Z_{H} (q^2 - \mH^2) - \de \mH^2\,, \\
\hSi_{hH}(q^2)
&= \Si_{hH}(q^2) + \edz \de Z_{Hh} (q^2 - \mH^2)
                 + \edz \de Z_{hH} (q^2 - \mh^2)
                 - \de \mhH^2\,, \\
\hSi_{hh}(q^2)
&= \Si_{hh}(q^2) + \de Z_{h} (q^2 - \mh^2)
                 - \de \mh^2\,.
\end{aligned}
\label{eq:renSihiggs}
\end{equation}
The mass counter terms arise from the renormalization of the Higgs
potential, see~\citere{MSSMren}. They are evaluated in the on-shell
renormalization scheme.  The field-renormalization constants can be
obtained in the \msbar\ scheme,%
\footnote{
Our results have been obtained using Dimensional Reduction
(DRED)~\cite{dred}.
}
leading to
\begin{equation}
\begin{aligned}
\de Z_{H} &= -\KKL \re\Si_{HH}'(\mH^2) \KKR^{\rm div}, \\
\de Z_{h} &= -\KKL \re\Si_{hh}'(\mh^2) \KKR^{\rm div} , \\
\de Z_{hH} &= \frac{\Sa\Ca}{\CZa} (\de Z_{h} - \de Z_{H})\,, \\
\de Z_{Hh} &= \de Z_{hH}\, ,
\end{aligned}
\label{eq:deltaZHiggs}
\end{equation}
i.e.\ only the divergent parts of the renormalization constants in
\refeqs{eq:deltaZHiggs} are taken into account.  As renormalization
scale we have chosen $\mu^{\msbarm} = \mt$.


\section{The process \eenenehH}
\label{sec:WWhH}

\subsection{The tree-level process}
\label{subsec:tree}

\begin{figure}[ht!]
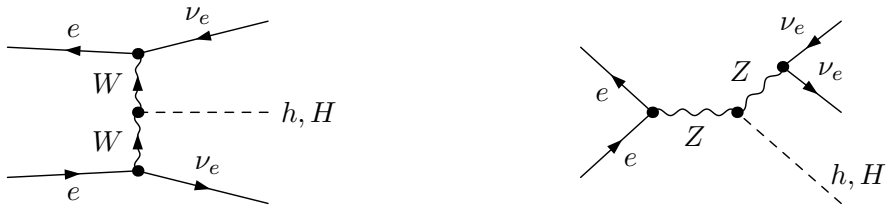

\begin{center}
\begin{small}
\unitlength=1.5bp%
\begin{feynartspicture}(216,72)(3,1)
\FADiagram{}
\FAProp(0.,15.)(10.,14.5)(0.,){/Straight}{-1}
\FALabel(5.0774,15.8181)[b]{$e$}
\FAProp(0.,5.)(10.,5.5)(0.,){/Straight}{1}
\FALabel(5.0774,4.18193)[t]{$e$}
\FAProp(20.,17.)(10.,14.5)(0.,){/Straight}{1}
\FALabel(14.6241,16.7737)[b]{$\nu_e$}
\FAProp(20.,10.)(10.,10.)(0.,){/ScalarDash}{0}
\FALabel(21.,8.82)[lb]{$h, H$}
\FAProp(20.,3.)(10.,5.5)(0.,){/Straight}{-1}
\FALabel(15.3759,5.27372)[b]{$\nu_e$}
\FAProp(10.,14.5)(10.,10.)(0.,){/Sine}{-1}
\FALabel(8.93,12.25)[r]{$W$}
\FAProp(10.,5.5)(10.,10.)(0.,){/Sine}{1}
\FALabel(8.93,7.75)[r]{$W$}
\FAVert(10.,14.5){0}
\FAVert(10.,5.5){0}
\FAVert(10.,10.){0}

\FADiagram{}

\FADiagram{}
\FAProp(0.,15.)(5.5,10.)(0.,){/Straight}{-1}
\FALabel(2.18736,11.8331)[tr]{$e$}
\FAProp(0.,5.)(5.5,10.)(0.,){/Straight}{1}
\FALabel(3.31264,6.83309)[tl]{$e$}
\FAProp(20.,17.)(15.5,13.5)(0.,){/Straight}{1}
\FALabel(17.2784,15.9935)[br]{$\nu_e$}
\FAProp(20.,10.)(15.5,13.5)(0.,){/Straight}{-1}
\FALabel(18.2216,12.4935)[bl]{$\nu_e$}
\FAProp(20.,3.)(12.,10.)(0.,){/ScalarDash}{0}
\FALabel(23.5,6.00165)[tr]{$h,H$}
\FAProp(5.5,10.)(12.,10.)(0.,){/Sine}{0}
\FALabel(8.75,8.93)[t]{$Z$}
\FAProp(15.5,13.5)(12.,10.)(0.,){/Sine}{0}
\FALabel(13.134,12.366)[br]{$Z$}
\FAVert(5.5,10.){0}
\FAVert(15.5,13.5){0}
\FAVert(12.,10.){0}
\end{feynartspicture}
\end{small}
\caption{%
The tree-level diagrams for the process \eennhH, consisting of the
$WW$-fusion contribution (left) and the Higgs-strahlung contribution
(right). 
}
\label{fig:tree}
\end{center}
\end{figure}

The tree-level process~\cite{eennHtree0,eennHtree}
consists of the two diagrams shown in
\reffi{fig:tree}. Besides the $WW$-fusion contribution (left diagram),
we also take into account the Higgs-strahlung contribution (right
diagram), where a virtual $Z$~boson is connected to two 
electron neutrinos.

An analytical expression for the tree-level cross section for an SM
Higgs boson can be found e.g.\ in \citere{eennHtree2}. 
For relatively low energies and moderate values of the SM Higgs-boson mass 
($\sqrt{s} \lesssim 400 \gev$, $\MHSM \lesssim 200 \gev$) the resonant
production via the Higgs-strahlung contribution dominates over the
$WW$-fusion contribution. 
At higher energies,
however, the $WW$-fusion contribution becomes dominant.
The cross section, containing both contributions, in the
high-energy limit takes the simple form~\cite{eennHtree}
\begin{equation}
\si(e^+ e^- \to \bar\nu_e\nu_e H_{\rm SM})\to
\frac{G_F^3 \MW^4}{4 \wz \, \pi^3} 
 \KKL \KL 1 + \frac{\MHSM^2}{s} \KR \log\KL\frac{s}{\MHSM^2}\KR
     - 2 \KL 1 - \frac{\MHSM^2}{s} \KR \KKR ,
\end{equation}
where the $t$-channel contribution from the $WW$-fusion diagram
gives rise to the logarithmic increase.
For our numerical results we use the full MSSM tree-level matrix
element, see \refse{subsec:HOXS}.

The coefficients for the couplings 
$WWh$ and $WWH$ are denoted by $\Ghn$ and $\GHn$ at the tree level,
respectively (and analogously for the $ZZh$ and $ZZH$ couplings): 
\begin{align}
\label{eq:WWhtree}
\Ghn &= \frac{\ri\,e\,\MW}{\sw} \Sba\,, \\
\label{eq:WWHtree}
\GHn &= \frac{\ri\,e\,\MW}{\sw} \Cba\,.
\end{align}
The SM coupling $\Ga^{(0)}_{H_{\rm SM}}$ is obtained by dropping the
SUSY factors $\Sba$ or $\Cba$.  In the decoupling limit, $\MA\gtrsim 200
\gev$, $\be - \al \to \pi/2$, so that $\Sba \to 1$ and $\Cba \to 0$, \ie
the heavy neutral $\cp$-even Higgs boson decouples from the $W$~and
$Z$~bosons. 

We parametrize the Born matrix element by the Fermi constant, $\gf$,
i.e.\ we use the relation 
\begin{equation}
e = 2 \; \sw \; \MW \KKL \frac{\wz \, \gf}
                              {1 + \De r} \KKR^{1/2} ,
\label{gmueparam}
\end{equation}
where $\De r$ incorporates higher-order corrections, see
\refse{subsec:higherorder}.


\subsection{Higher-order corrections}
\label{subsec:higherorder}

In the description of our calculation below we will mainly concentrate 
on the $WW$-fusion contribution.  
The Higgs-strahlung contribution, which we describe in less 
detail, is taken into account in exactly the same way (see, however,
\refse{subsec:HOXS}). 

We evaluate the \onel\ \order{\al} contributions from loops involving
all fermions and sfermions.
Especially the corrections involving third-generation fermions and 
sfermions, i.e.\ 
$t, b, \tau, \nu_\tau$, and their corresponding superpartners, 
$\Stope, \Stopz$, $\Sbote, \Sbotz$, $\Staue, \Stauz$, $\Sneut$, are
expected to be sizable, 
since they contain potentially large Yukawa couplings, 
$y_t$, $y_b$, $y_\tau$, where the down-type couplings can be enhanced 
in the MSSM for large values of $\tb$. This class of diagrams
in particular contains contributions enhanced by $\mt^2/\MW^2$. 

\begin{figure}[ht!]
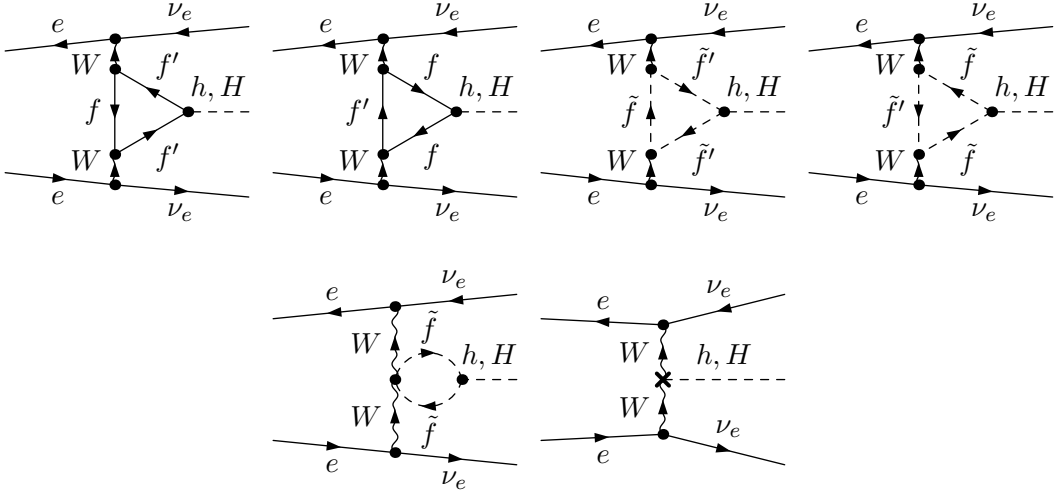

\vspace{2em}
\begin{center}
\begin{small}
\unitlength=1bp%
\begin{feynartspicture}(432,202)(4,2)
\FADiagram{}
\FAProp(0.,15.)(9.,16.)(0.,){/Straight}{-1}
\FALabel(4.32883,16.5605)[b]{$e$}
\FAProp(0.,5.)(9.,4.)(0.,){/Straight}{1}
\FALabel(4.32883,3.43948)[t]{$e$}
\FAProp(20.,17.)(9.,16.)(0.,){/Straight}{1}
\FALabel(14.3597,17.5636)[b]{$\nu_e$}
\FAProp(20.,10.)(15.,10.)(0.,){/ScalarDash}{0}
\FALabel(17.5,10.82)[b]{$h, H$}
\FAProp(20.,3.)(9.,4.)(0.,){/Straight}{-1}
\FALabel(14.3597,2.43637)[t]{$\nu_e$}
\FAProp(9.,16.)(9.,13.5)(0.,){/Sine}{-1}
\FALabel(7.93,14.75)[tr]{$W$}
\FAProp(9.,4.)(9.,6.5)(0.,){/Sine}{1}
\FALabel(7.93,5.25)[br]{$W$}
\FAProp(15.,10.)(9.,13.5)(0.,){/Straight}{1}
\FALabel(12.301,12.6089)[bl]{$f'$}
\FAProp(15.,10.)(9.,6.5)(0.,){/Straight}{-1}
\FALabel(12.301,7.39114)[tl]{$f'$}
\FAProp(9.,13.5)(9.,6.5)(0.,){/Straight}{1}
\FALabel(7.93,10.)[r]{$f$}
\FAVert(9.,16.){0}
\FAVert(9.,4.){0}
\FAVert(15.,10.){0}
\FAVert(9.,13.5){0}
\FAVert(9.,6.5){0}

\FADiagram{}
\FAProp(0.,15.)(9.,16.)(0.,){/Straight}{-1}
\FALabel(4.32883,16.5605)[b]{$e$}
\FAProp(0.,5.)(9.,4.)(0.,){/Straight}{1}
\FALabel(4.32883,3.43948)[t]{$e$}
\FAProp(20.,17.)(9.,16.)(0.,){/Straight}{1}
\FALabel(14.3597,17.5636)[b]{$\nu_e$}
\FAProp(20.,10.)(15.,10.)(0.,){/ScalarDash}{0}
\FALabel(17.5,10.82)[b]{$h, H$}
\FAProp(20.,3.)(9.,4.)(0.,){/Straight}{-1}
\FALabel(14.3597,2.43637)[t]{$\nu_e$}
\FAProp(9.,16.)(9.,13.5)(0.,){/Sine}{-1}
\FALabel(7.93,14.75)[tr]{$W$}
\FAProp(9.,4.)(9.,6.5)(0.,){/Sine}{1}
\FALabel(7.93,5.25)[br]{$W$}
\FAProp(15.,10.)(9.,13.5)(0.,){/Straight}{-1}
\FALabel(12.301,12.6089)[bl]{$f$}
\FAProp(15.,10.)(9.,6.5)(0.,){/Straight}{1}
\FALabel(12.301,7.39114)[tl]{$f$}
\FAProp(9.,13.5)(9.,6.5)(0.,){/Straight}{-1}
\FALabel(7.93,10.)[r]{$f'$}
\FAVert(9.,16.){0}
\FAVert(9.,4.){0}
\FAVert(15.,10.){0}
\FAVert(9.,13.5){0}
\FAVert(9.,6.5){0}

\FADiagram{}
\FAProp(0.,15.)(9.,16.)(0.,){/Straight}{-1}
\FALabel(4.32883,16.5605)[b]{$e$}
\FAProp(0.,5.)(9.,4.)(0.,){/Straight}{1}
\FALabel(4.32883,3.43948)[t]{$e$}
\FAProp(20.,17.)(9.,16.)(0.,){/Straight}{1}
\FALabel(14.3597,17.5636)[b]{$\nu_e$}
\FAProp(20.,10.)(15.,10.)(0.,){/ScalarDash}{0}
\FALabel(17.5,10.82)[b]{$h, H$}
\FAProp(20.,3.)(9.,4.)(0.,){/Straight}{-1}
\FALabel(14.3597,2.43637)[t]{$\nu_e$}
\FAProp(9.,16.)(9.,13.5)(0.,){/Sine}{-1}
\FALabel(7.93,14.75)[tr]{$W$}
\FAProp(9.,4.)(9.,6.5)(0.,){/Sine}{1}
\FALabel(7.93,5.25)[br]{$W$}
\FAProp(15.,10.)(9.,13.5)(0.,){/ScalarDash}{-1}
\FALabel(12.301,12.6089)[bl]{$\tilde f'$}
\FAProp(15.,10.)(9.,6.5)(0.,){/ScalarDash}{1}
\FALabel(12.301,7.39114)[tl]{$\tilde f'$}
\FAProp(9.,13.5)(9.,6.5)(0.,){/ScalarDash}{-1}
\FALabel(7.93,10.)[r]{$\tilde f$}
\FAVert(9.,16.){0}
\FAVert(9.,4.){0}
\FAVert(15.,10.){0}
\FAVert(9.,13.5){0}
\FAVert(9.,6.5){0}

\FADiagram{}
\FAProp(0.,15.)(9.,16.)(0.,){/Straight}{-1}
\FALabel(4.32883,16.5605)[b]{$e$}
\FAProp(0.,5.)(9.,4.)(0.,){/Straight}{1}
\FALabel(4.32883,3.43948)[t]{$e$}
\FAProp(20.,17.)(9.,16.)(0.,){/Straight}{1}
\FALabel(14.3597,17.5636)[b]{$\nu_e$}
\FAProp(20.,10.)(15.,10.)(0.,){/ScalarDash}{0}
\FALabel(17.5,10.82)[b]{$h, H$}
\FAProp(20.,3.)(9.,4.)(0.,){/Straight}{-1}
\FALabel(14.3597,2.43637)[t]{$\nu_e$}
\FAProp(9.,16.)(9.,13.5)(0.,){/Sine}{-1}
\FALabel(7.93,14.75)[tr]{$W$}
\FAProp(9.,4.)(9.,6.5)(0.,){/Sine}{1}
\FALabel(7.93,5.25)[br]{$W$}
\FAProp(15.,10.)(9.,13.5)(0.,){/ScalarDash}{1}
\FALabel(12.301,12.6089)[bl]{$\tilde f$}
\FAProp(15.,10.)(9.,6.5)(0.,){/ScalarDash}{-1}
\FALabel(12.301,7.39114)[tl]{$\tilde f$}
\FAProp(9.,13.5)(9.,6.5)(0.,){/ScalarDash}{1}
\FALabel(7.93,10.)[r]{$\tilde f'$}
\FAVert(9.,16.){0}
\FAVert(9.,4.){0}
\FAVert(15.,10.){0}
\FAVert(9.,13.5){0}
\FAVert(9.,6.5){0}

\FADiagram{}

\FADiagram{}
\FAProp(0.,15.)(10.,16.)(0.,){/Straight}{-1}
\FALabel(4.84577,16.5623)[b]{$e$}
\FAProp(0.,5.)(10.,4.)(0.,){/Straight}{1}
\FALabel(4.84577,3.43769)[t]{$e$}
\FAProp(20.,17.)(10.,16.)(0.,){/Straight}{1}
\FALabel(14.8458,17.5623)[b]{$\nu_e$}
\FAProp(20.,10.)(15.5,10.)(0.,){/ScalarDash}{0}
\FALabel(17.75,10.82)[b]{$h, H$}
\FAProp(20.,3.)(10.,4.)(0.,){/Straight}{-1}
\FALabel(14.8458,2.43769)[t]{$\nu_e$}
\FAProp(10.,16.)(10.,10.)(0.,){/Sine}{-1}
\FALabel(8.93,13.)[r]{$W$}
\FAProp(10.,4.)(10.,10.)(0.,){/Sine}{1}
\FALabel(8.93,7.)[r]{$W$}
\FAProp(15.5,10.)(10.,10.)(0.8,){/ScalarDash}{-1}
\FALabel(12.75,13)[b]{$\tilde f$}
\FAProp(15.5,10.)(10.,10.)(-0.8,){/ScalarDash}{1}
\FALabel(12.75,7)[t]{$\tilde f$}
\FAVert(10.,16.){0}
\FAVert(10.,4.){0}
\FAVert(15.5,10.){0}
\FAVert(10.,10.){0}

\FADiagram{}
\FAProp(0.,15.)(10.,14.5)(0.,){/Straight}{-1}
\FALabel(5.0774,15.8181)[b]{$e$}
\FAProp(0.,5.)(10.,5.5)(0.,){/Straight}{1}
\FALabel(5.0774,4.18193)[t]{$e$}
\FAProp(20.,17.)(10.,14.5)(0.,){/Straight}{1}
\FALabel(14.6241,16.7737)[b]{$\nu_e$}
\FAProp(20.,10.)(10.,10.)(0.,){/ScalarDash}{0}
\FALabel(15.,10.82)[b]{$h, H$}
\FAProp(20.,3.)(10.,5.5)(0.,){/Straight}{-1}
\FALabel(15.3759,5.27372)[b]{$\nu_e$}
\FAProp(10.,14.5)(10.,10.)(0.,){/Sine}{-1}
\FALabel(8.93,12.25)[r]{$W$}
\FAProp(10.,5.5)(10.,10.)(0.,){/Sine}{1}
\FALabel(8.93,7.75)[r]{$W$}
\FAVert(10.,14.5){0}
\FAVert(10.,5.5){0}
\FAVert(10.,10.){1}
\end{feynartspicture}
\end{small}
\vspace{1em}
\caption{
Corrections to the $WW\{h,H\}$ vertex and the corresponding counter-term
diagram.  The label $\xtilde f$ denotes all (s)fermions, except in the
presence of a $\xtilde f\,'$, in which case the former denotes only the
isospin-up and the latter the isospin-down members of the (s)fermion
doublets.
}
\label{fig:WWhHvert}
\end{center}
\end{figure}

\begin{figure}[ht!]
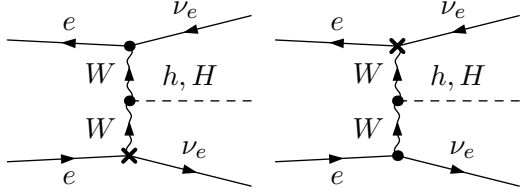

\begin{center}
\begin{small}
\unitlength=1bp%
\begin{feynartspicture}(216,101)(2,1)
\FADiagram{}
\FAProp(0.,15.)(10.,14.5)(0.,){/Straight}{-1}
\FALabel(5.0774,15.8181)[b]{$e$}
\FAProp(0.,5.)(10.,5.5)(0.,){/Straight}{1}
\FALabel(5.0774,4.18193)[t]{$e$}
\FAProp(20.,17.)(10.,14.5)(0.,){/Straight}{1}
\FALabel(14.6241,16.7737)[b]{$\nu_e$}
\FAProp(20.,10.)(10.,10.)(0.,){/ScalarDash}{0}
\FALabel(15.,10.82)[b]{$h, H$}
\FAProp(20.,3.)(10.,5.5)(0.,){/Straight}{-1}
\FALabel(15.3759,5.27372)[b]{$\nu_e$}
\FAProp(10.,14.5)(10.,10.)(0.,){/Sine}{-1}
\FALabel(8.93,12.25)[r]{$W$}
\FAProp(10.,5.5)(10.,10.)(0.,){/Sine}{1}
\FALabel(8.93,7.75)[r]{$W$}
\FAVert(10.,14.5){0}
\FAVert(10.,10.){0}
\FAVert(10.,5.5){1}

\FADiagram{}
\FAProp(0.,15.)(10.,14.5)(0.,){/Straight}{-1}
\FALabel(5.0774,15.8181)[b]{$e$}
\FAProp(0.,5.)(10.,5.5)(0.,){/Straight}{1}
\FALabel(5.0774,4.18193)[t]{$e$}
\FAProp(20.,17.)(10.,14.5)(0.,){/Straight}{1}
\FALabel(14.6241,16.7737)[b]{$\nu_e$}
\FAProp(20.,10.)(10.,10.)(0.,){/ScalarDash}{0}
\FALabel(15.,10.82)[b]{$h, H$}
\FAProp(20.,3.)(10.,5.5)(0.,){/Straight}{-1}
\FALabel(15.3759,5.27372)[b]{$\nu_e$}
\FAProp(10.,14.5)(10.,10.)(0.,){/Sine}{-1}
\FALabel(8.93,12.25)[r]{$W$}
\FAProp(10.,5.5)(10.,10.)(0.,){/Sine}{1}
\FALabel(8.93,7.75)[r]{$W$}
\FAVert(10.,5.5){0}
\FAVert(10.,10.){0}
\FAVert(10.,14.5){1}
\end{feynartspicture}
\end{small}
\caption{
Counter-term contributions entering via the $e\,\nu_e\,W$ vertex.
}
\label{fig:enuWCT}
\end{center}
\end{figure}

The contributions involve corrections to the $WW\{h,H\}$ vertex and the
corresponding counter-term diagram, shown in \reffi{fig:WWhHvert},
corrections to the $W$-boson propagators and the corresponding counter
terms, shown in \reffi{fig:WWhHself}, and the counter-term contributions
to the $e\nu_e W$ vertex as shown in \reffi{fig:enuWCT}. Furthermore,
Higgs propagator corrections enter via the wave-function normalization of
the external Higgs boson, see \refse{subsec:higgsprop} below. There are
also $W$-boson propagator corrections inducing a transition from the
$W^\pm$ to either $G^\pm$ or $H^\pm$.  These corrections affect only the
longitudinal part of the $W$~boson, however, and are thus $\propto
m_e/\MW$ and have been neglected.

\begin{figure}[ht!]
\vspace{2em}
\begin{center}
\begin{small}
\unitlength=1bp%
\begin{feynartspicture}(432,202)(4,2)
\FADiagram{}
\FAProp(0.,15.)(10.,16.5)(0.,){/Straight}{-1}
\FALabel(4.77007,16.8029)[b]{$e$}
\FAProp(0.,5.)(10.,3.5)(0.,){/Straight}{1}
\FALabel(4.77007,3.19715)[t]{$e$}
\FAProp(20.,17.)(10.,16.5)(0.,){/Straight}{1}
\FALabel(14.9226,17.8181)[b]{$\nu_e$}
\FAProp(20.,10.)(10.,13.5)(0.,){/ScalarDash}{0}
\FALabel(15.3153,12.17)[lb]{$h, H$}
\FAProp(20.,3.)(10.,3.5)(0.,){/Straight}{-1}
\FALabel(14.9226,2.18193)[t]{$\nu_e$}
\FAProp(10.,16.5)(10.,13.5)(0.,){/Sine}{-1}
\FALabel(11.07,15.)[l]{$W$}
\FAProp(10.,3.5)(10.,6.)(0.,){/Sine}{1}
\FALabel(11.07,4.75)[l]{$W$}
\FAProp(10.,13.5)(10.,11.)(0.,){/Sine}{-1}
\FALabel(8.93,12.25)[r]{$W$}
\FAProp(10.,6.)(10.,11.)(0.8,){/Straight}{-1}
\FALabel(13.07,8.5)[l]{$f$}
\FAProp(10.,6.)(10.,11.)(-0.8,){/Straight}{1}
\FALabel(6.93,8.5)[r]{$f'$}
\FAVert(10.,16.5){0}
\FAVert(10.,3.5){0}
\FAVert(10.,13.5){0}
\FAVert(10.,6.){0}
\FAVert(10.,11.){0}

\FADiagram{}
\FAProp(0.,15.)(10.,16.5)(0.,){/Straight}{-1}
\FALabel(4.77007,16.8029)[b]{$e$}
\FAProp(0.,5.)(10.,3.5)(0.,){/Straight}{1}
\FALabel(4.77007,3.19715)[t]{$e$}
\FAProp(20.,17.)(10.,16.5)(0.,){/Straight}{1}
\FALabel(14.9226,17.8181)[b]{$\nu_e$}
\FAProp(20.,10.)(10.,13.5)(0.,){/ScalarDash}{0}
\FALabel(15.3153,12.17)[lb]{$h, H$}
\FAProp(20.,3.)(10.,3.5)(0.,){/Straight}{-1}
\FALabel(14.9226,2.18193)[t]{$\nu_e$}
\FAProp(10.,16.5)(10.,13.5)(0.,){/Sine}{-1}
\FALabel(11.07,15.)[l]{$W$}
\FAProp(10.,3.5)(10.,6.)(0.,){/Sine}{1}
\FALabel(11.07,4.75)[l]{$W$}
\FAProp(10.,13.5)(10.,11.)(0.,){/Sine}{-1}
\FALabel(8.93,12.25)[r]{$W$}
\FAProp(10.,6.)(10.,11.)(0.8,){/ScalarDash}{-1}
\FALabel(13.07,8.5)[l]{$\tilde f$}
\FAProp(10.,6.)(10.,11.)(-0.8,){/ScalarDash}{1}
\FALabel(6.93,8.5)[r]{$\tilde f'$}
\FAVert(10.,16.5){0}
\FAVert(10.,3.5){0}
\FAVert(10.,13.5){0}
\FAVert(10.,6.){0}
\FAVert(10.,11.){0}

\FADiagram{}
\FAProp(0.,15.)(10.,16.)(0.,){/Straight}{-1}
\FALabel(4.84577,16.5623)[b]{$e$}
\FAProp(0.,5.)(10.,4.)(0.,){/Straight}{1}
\FALabel(4.84577,3.43769)[t]{$e$}
\FAProp(20.,17.)(10.,16.)(0.,){/Straight}{1}
\FALabel(14.8458,17.5623)[b]{$\nu_e$}
\FAProp(20.,10.)(10.,12.5)(0.,){/ScalarDash}{0}
\FALabel(15.3153,12.0312)[lb]{$h, H$}
\FAProp(20.,3.)(10.,4.)(0.,){/Straight}{-1}
\FALabel(14.8458,2.43769)[t]{$\nu_e$}
\FAProp(10.,16.)(10.,12.5)(0.,){/Sine}{-1}
\FALabel(8.93,14.25)[r]{$W$}
\FAProp(10.,4.)(10.,8.5)(0.,){/Sine}{1}
\FALabel(11.07,6.25)[l]{$W$}
\FAProp(10.,12.5)(10.,8.5)(0.,){/Sine}{-1}
\FALabel(11.12,10.15)[l]{$W$}
\FAProp(10.,8.5)(10.,8.5)(5.5,8.5){/ScalarDash}{-1}
\FALabel(4.43,8.5)[r]{$\tilde f$}
\FAVert(10.,16.){0}
\FAVert(10.,4.){0}
\FAVert(10.,12.5){0}
\FAVert(10.,8.5){0}

\FADiagram{}
\FAProp(0.,15.)(10.,15.5)(0.,){/Straight}{-1}
\FALabel(4.9226,16.3181)[b]{$e$}
\FAProp(0.,5.)(10.,4.5)(0.,){/Straight}{1}
\FALabel(4.9226,3.68193)[t]{$e$}
\FAProp(20.,17.)(10.,15.5)(0.,){/Straight}{1}
\FALabel(14.7701,17.3029)[b]{$\nu_e$}
\FAProp(20.,10.)(10.,11.5)(0.,){/ScalarDash}{0}
\FALabel(15.3153,11.5556)[lb]{$h, H$}
\FAProp(20.,3.)(10.,4.5)(0.,){/Straight}{-1}
\FALabel(15.2299,4.80285)[b]{$\nu_e$}
\FAProp(10.,8.)(10.,4.5)(0.,){/Sine}{-1}
\FALabel(8.93,6.25)[r]{$W$}
\FAProp(10.,8.)(10.,11.5)(0.,){/Sine}{1}
\FALabel(8.93,9.75)[r]{$W$}
\FAProp(10.,15.5)(10.,11.5)(0.,){/Sine}{-1}
\FALabel(8.93,13.5)[r]{$W$}
\FAVert(10.,15.5){0}
\FAVert(10.,4.5){0}
\FAVert(10.,11.5){0}
\FAVert(10.,8.){1}

\FADiagram{}
\FAProp(0.,15.)(10.,16.5)(0.,){/Straight}{-1}
\FALabel(4.77007,16.8029)[b]{$e$}
\FAProp(0.,5.)(10.,3.5)(0.,){/Straight}{1}
\FALabel(4.77007,3.19715)[t]{$e$}
\FAProp(20.,17.)(10.,16.5)(0.,){/Straight}{1}
\FALabel(14.9226,17.8181)[b]{$\nu_e$}
\FAProp(20.,10.)(10.,6.5)(0.,){/ScalarDash}{0}
\FALabel(15.3153,7.68)[lt]{$h, H$}
\FAProp(20.,3.)(10.,3.5)(0.,){/Straight}{-1}
\FALabel(14.9226,2.18193)[t]{$\nu_e$}
\FAProp(10.,16.5)(10.,14.)(0.,){/Sine}{-1}
\FALabel(11.07,15.25)[l]{$W$}
\FAProp(10.,3.5)(10.,6.5)(0.,){/Sine}{1}
\FALabel(11.07,5.)[l]{$W$}
\FAProp(10.,6.5)(10.,9.)(0.,){/Sine}{1}
\FALabel(8.93,7.75)[r]{$W$}
\FAProp(10.,14.)(10.,9.)(0.8,){/Straight}{1}
\FALabel(6.93,11.5)[r]{$f$}
\FAProp(10.,14.)(10.,9.)(-0.8,){/Straight}{-1}
\FALabel(13.07,11.5)[l]{$f'$}
\FAVert(10.,16.5){0}
\FAVert(10.,3.5){0}
\FAVert(10.,6.5){0}
\FAVert(10.,14.){0}
\FAVert(10.,9.){0}

\FADiagram{}
\FAProp(0.,15.)(10.,16.5)(0.,){/Straight}{-1}
\FALabel(4.77007,16.8029)[b]{$e$}
\FAProp(0.,5.)(10.,3.5)(0.,){/Straight}{1}
\FALabel(4.77007,3.19715)[t]{$e$}
\FAProp(20.,17.)(10.,16.5)(0.,){/Straight}{1}
\FALabel(14.9226,17.8181)[b]{$\nu_e$}
\FAProp(20.,10.)(10.,6.5)(0.,){/ScalarDash}{0}
\FALabel(15.3153,7.68)[lt]{$h, H$}
\FAProp(20.,3.)(10.,3.5)(0.,){/Straight}{-1}
\FALabel(14.9226,2.18193)[t]{$\nu_e$}
\FAProp(10.,16.5)(10.,14.)(0.,){/Sine}{-1}
\FALabel(11.07,15.25)[l]{$W$}
\FAProp(10.,3.5)(10.,6.5)(0.,){/Sine}{1}
\FALabel(11.07,5.)[l]{$W$}
\FAProp(10.,6.5)(10.,9.)(0.,){/Sine}{1}
\FALabel(8.93,7.75)[r]{$W$}
\FAProp(10.,14.)(10.,9.)(0.8,){/ScalarDash}{1}
\FALabel(6.93,11.5)[r]{$\tilde f$}
\FAProp(10.,14.)(10.,9.)(-0.8,){/ScalarDash}{-1}
\FALabel(13.07,11.5)[l]{$\tilde f'$}
\FAVert(10.,16.5){0}
\FAVert(10.,3.5){0}
\FAVert(10.,6.5){0}
\FAVert(10.,14.){0}
\FAVert(10.,9.){0}

\FADiagram{}
\FAProp(0.,15.)(10.,16.)(0.,){/Straight}{-1}
\FALabel(4.84577,16.5623)[b]{$e$}
\FAProp(0.,5.)(10.,4.)(0.,){/Straight}{1}
\FALabel(4.84577,3.43769)[t]{$e$}
\FAProp(20.,17.)(10.,16.)(0.,){/Straight}{1}
\FALabel(14.8458,17.5623)[b]{$\nu_e$}
\FAProp(20.,10.)(10.,8.)(0.,){/ScalarDash}{0}
\FALabel(15.3153,8.20525)[lt]{$h, H$}
\FAProp(20.,3.)(10.,4.)(0.,){/Straight}{-1}
\FALabel(14.8458,2.43769)[t]{$\nu_e$}
\FAProp(10.,16.)(10.,11.5)(0.,){/Sine}{-1}
\FALabel(11.07,13.75)[l]{$W$}
\FAProp(10.,4.)(10.,8.)(0.,){/Sine}{1}
\FALabel(8.93,6.)[r]{$W$}
\FAProp(10.,8.)(10.,11.5)(0.,){/Sine}{1}
\FALabel(11.12,10.1)[l]{$W$}
\FAProp(10.,11.5)(10.,11.5)(5.5,11.5){/ScalarDash}{-1}
\FALabel(4.43,11.5)[r]{$\tilde f$}
\FAVert(10.,16.){0}
\FAVert(10.,4.){0}
\FAVert(10.,8.){0}
\FAVert(10.,11.5){0}

\FADiagram{}
\FAProp(0.,15.)(10.,15.5)(0.,){/Straight}{-1}
\FALabel(5.0774,14.1819)[t]{$e$}
\FAProp(0.,5.)(10.,4.5)(0.,){/Straight}{1}
\FALabel(4.9226,3.68193)[t]{$e$}
\FAProp(20.,17.)(10.,15.5)(0.,){/Straight}{1}
\FALabel(14.7701,17.3029)[b]{$\nu_e$}
\FAProp(20.,10.)(10.,8.5)(0.,){/ScalarDash}{0}
\FALabel(15.3153,10.556)[lb]{$h, H$}
\FAProp(20.,3.)(10.,4.5)(0.,){/Straight}{-1}
\FALabel(14.7701,2.69715)[t]{$\nu_e$}
\FAProp(10.,12.)(10.,15.5)(0.,){/Sine}{1}
\FALabel(11.07,13.75)[l]{$W$}
\FAProp(10.,12.)(10.,8.5)(0.,){/Sine}{-1}
\FALabel(8.93,10.25)[r]{$W$}
\FAProp(10.,4.5)(10.,8.5)(0.,){/Sine}{1}
\FALabel(11.07,6.5)[l]{$W$}
\FAVert(10.,15.5){0}
\FAVert(10.,4.5){0}
\FAVert(10.,8.5){0}
\FAVert(10.,12.){1}
\end{feynartspicture}
\end{small}
\vspace{1em}
\caption{
Corrections to the $W$-boson propagator and the corresponding counter-term
diagrams.  The label $\xtilde f$ denotes all (s)fermions, except
in the presence of a $\xtilde f\,'$, in which case the former denotes only
the isospin-up and the latter the isospin-down members of the
(s)fermion doublets.
}
\label{fig:WWhHself}
\end{center}
\end{figure}

While the renormalization in the counter terms depicted in
\reffis{fig:WWhHself} and \ref{fig:enuWCT} is as in the SM (see e.g.\
\citere{ansgarhabil}), the
$WW\{h,H\}$ vertices are renormalized as follows,
\begin{align}
WWh:\quad
\Ga_{WWh}^{(0), {\rm CT}} = \Ghn & 
  \Biggl[ 1 + \de \tilde Z_e 
            + \edz \frac{\de \MW^2}{\MW^2}
            + \de Z_W 
            + \frac{\de \sw}{\sw} \notag \\
&{}
            + \sinb\Cb \frac{\Cba}{\Sba} \de\tb
            + \edz \de Z_h
            + \edz \frac{\GHn}{\Ghn} \de Z_{Hh}
  \Biggr]\,,
\label{eq:WWhct} \\
WWH:\quad
\Ga_{WWH}^{(0), {\rm CT}} = \GHn & 
  \Biggl[ 1 + \de \tilde Z_e 
            + \edz \frac{\de \MW^2}{\MW^2}
            + \de Z_W 
            + \frac{\de \sw}{\sw} \notag \\
&{}
            - \sinb\Cb \frac{\Sba}{\Cba} \de \tb
            + \edz \de Z_H
            + \edz \frac{\Ghn}{\GHn} \de Z_{hH}
  \Biggr]\,.
\label{eq:WWHct}
\end{align}
Analogous expressions are obtained for $\Ga_{ZZ\Phi}^{(0),{\rm CT}}$ 
($\Phi = h,H$). In the above expressions
$\de \tilde Z_e$ incorporates the charge renormalization and the
$\De r$ contribution arising from \refeq{gmueparam}, 
\begin{equation}
\de \tilde Z_e = \de Z_e - \edz \De r, \quad
\de Z_e = \edz\Pi^\ga(0) - \frac{\sw}{\cw} 
                           \frac{\Si_{\ga Z}^T(0)}{\MZ^2}\,,
\end{equation}
where $\Si^T$ denotes the transverse part of a self-energy.
$\de\MW^2$ is
the $W$-mass counter term, $\de Z_W$ is the corresponding
field-renormalization constant, and $\de\sw$ denotes the renormalization 
constant for
the weak mixing angle. The field-renormalization constants, $\de
Z_H, \de Z_h$, and $\de Z_{Hh} = \de Z_{hH}$ are given in
\refeq{eq:deltaZHiggs}.  The counter term for $\tb$ (with $\tb \to \tb
(1 + \de \tb)$) is derived in the \msbar\ renormalization
scheme~\cite{MSSMren} (using DRED). The parameter $\tb$ in our result thus
corresponds to the \msbar\ parameter, taken at the scale 
$\mu^{\msbarm} = \mt$.
We list here all contributing
counter terms except for the Higgs field renormalization which has already
been given in \refeq{eq:deltaZHiggs}:
\begin{equation}
\begin{aligned}
\de\MW^2 &= \re\Si_W^T(\MW^2)\,, \\[.5em]
\de\MZ^2 &= \re\Si_Z^T(\MZ^2)\,, \\[.5em]
\frac{\de\sw}{\sw} &= \edz \frac{\cw^2}{\sw^2}
  \KL \frac{\de\MZ^2}{\MZ^2} - \frac{\de\MW^2}{\MW^2} \KR, \\
\de\tb &= \de\tb^{\msbarm} = - \ed{2\CZa} 
  \KKL \re\Si_{hh}'(\mh^2) - \re\Si_{HH}'(\mH^2) \KKR^{\rm div}.
\end{aligned}
\label{eq:rcs}
\end{equation}
For $\de \tilde Z_e$ we find, taking into account only contributions
from fermion and sfermion loops,
\begin{equation}
\de \tilde Z_e = \edz \KKKL \frac{\cw^2}{\sw^2} 
                \KL \frac{\de\MZ^2}{\MZ^2} - \frac{\de\MW^2}{\MW^2} \KR
                - \KKL \frac{\Si_W^T(0) - \de\MW^2}{\MW^2} \KKR 
                \KKKR \,.
\end{equation}
The gauge-boson field-renormalization constants,
$\de Z_W, \de Z_Z, \de Z_{\ga Z}$, drop out in the result for the
complete $S$-matrix element. 

In order to ensure the correct on-shell properties of the outgoing
Higgs boson, which are necessary for the correct renormalization of 
the $S$-matrix element, furthermore finite wave-function normalizations 
have to be incorporated, see \refse{subsec:higgsprop} below.

We have performed several checks of the described renormalization
procedure. In particular we have verified that
\begin{itemize}
\item 
the $e\,\nu_e\,W$ vertex counter term is finite by itself,
\item
the gauge-boson field renormalizations drop out if all counter-term
diagrams are added up,
\item
the self-energy corrections together with their corresponding
counter-term diagrams are finite,
\item
the complete set of diagrams together with all counter-term diagrams
yields a finite result.
\end{itemize}

The individual contributions from fermions and sfermions constitute two
subsets of the full result which are individually UV-finite. This is in
contrast to the evaluation of renormalized Higgs boson self-energies
(see e.g.\ \citere{mhiggsf1l}), where a UV-finite result is obtained
only after adding the fermion- and sfermion-loop contributions.


\subsection{The Higgs-boson propagator corrections and the effective Born
  approximation}
\label{subsec:higgsprop}

For the correct normalization of the $S$-matrix element, finite
Higgs-boson propagator corrections have to be included such that the
residues of the outgoing Higgs bosons are set to unity and no mixing
between $h$ and $H$ occurs on the mass shell of the two particles. 
The corrections affecting the Higgs-boson propagators and the Higgs-boson
masses are numerically very important. Therefore we go beyond the
\onel\ fermion/sfermion contribution used for the evaluation of the
genuine \onel\ diagrams and include Higgs-boson corrections also from
other sectors of the model~\cite{mhiggsf1l} as well as the dominant
\twol\ contributions~\cite{mhiggsletter,mhiggslong,mhiggsEP3} as
incorporated in the program \fh~\cite{feynhiggs}.

For the $WW\{h,H\}$ vertex, these contributions can be
included as follows, yielding the correct normalization of the
$S$ matrix at \onel\ order%
\footnote{
Note that our notation is slightly different from
\citeres{hff,eehZhA}. 
}%
:
\begin{equation}
\begin{aligned}
\label{eq:ewdecayamplitude}
WWh &: \quad \wrZh \KL \Ghn + \edz \rZHh\; \GHn \KR, \\
WWH &: \quad \wrZH \KL \GHn + \edz \rZhH\; \Ghn \KR\,.
\end{aligned}
\end{equation}
This gives rise to the following terms: 
\begin{align}
WWh:\quad
\Ga_{WWh}^{{\rm WF}} = \Ghn & \;
  \Biggl[   \KL \sqrt{\rZh} - 1 \KR 
                 + \edz \frac{\GHn}{\Ghn} \sqrt{\rZh} \; \rZHh 
  \Biggr]\,,
\label{eq:WWhct2} \\
WWH:\quad
\Ga_{WWH}^{{\rm WF}} = \GHn & \;
  \Biggl[ \KL \sqrt{\rZH} - 1 \KR 
                 + \edz \frac{\Ghn}{\GHn} \sqrt{\rZH} \; \rZhH 
  \Biggr]\,.
\label{eq:WWHct2}
\end{align}

\noindent
Analogous expressions are obtained for $\Ga_{ZZ\Phi}^{\rm WF}$ 
($\Phi = h,H$).
In the above expressions, the finite Higgs-mixing contributions enter,
\begin{align}
\label{eq:ZhH}
\rZHh &= -2 \; \frac{\re\hSihH(\Mh^2)}{\Mh^2 - \mH^2 + \re\hSiH(\Mh^2)}\,, \\
\label{eq:ZHh}
\rZhH &= -2 \; \frac{\re\hSihH(\MH^2)}{\MH^2 - \mh^2 + \re\hSih(\MH^2)}\,,
\end{align}
involving the renormalized self-energies $\hSi(q^2)$, see
\refeq{eq:renSihiggs}, which contain corrections up to the \twol\ level.
The wave-function normalization factors $\rZh, \rZH$ are
related to the finite residue of the Higgs-boson propagators:
\begin{align}
\label{eq:zlh}
\rZh &= \ed{1 + \re\hSiph(q^2) - 
        \KL \frac{\KL \re\hSihH(q^2) \KR^2}
            {q^2 - \mH^2 + \re\hSiH(q^2)} \KR '}~_{\Bigr| q^2 = \Mh^2}\,, 
                                                                 \\[.5em]
\label{eq:zhh}
\rZH &= \ed{1 + \re\hSipH(q^2) -
        \KL \frac{\KL \re\hSihH(q^2) \KR^2}
            {q^2 - \mh^2 + \re\hSih(q^2)} \KR '}~_{\Bigr| q^2 = \MH^2}\,.
\end{align}

If in \refeqs{eq:ZhH}--(\ref{eq:zhh}) the renormalized self-energies
were evaluated at $q^2 = 0$, the above wave-function correction would
reduce to the $\aeff$~approximation~\cite{hff,eehZhA}. In this
approximation, however, the outgoing Higgs boson does not have the
correct on-shell properties. 

In order to analyze the effect of those corrections that go beyond the
universal Higgs propagator corrections, we 
include the Higgs propagator corrections according to
\refeqs{eq:WWhct2}--(\ref{eq:zhh}) into our Born matrix element, see
\refse{subsec:HOXS} below. 
Concerning our numerical analysis, see \refse{sec:numeval}, we either
use this Born cross section
(thus the difference between our tree-level
and the one-loop cross sections indicates the effect of the new genuine
loop corrections), or we use the $\aeff$~approximation (so that the
difference between the tree-level and the \onel\ cross section
directly shows the effect of our new calculation compared to the
previously used results).


\subsection{The higher-order production cross section}
\label{subsec:HOXS}

The amplitude for the process \eenenehH\ is denoted as
\begin{equation}
\cM^{(i)}_{\Phi,e}, \quad (\Phi = h,H;\ i = 0,1)\,,
\end{equation}
where $i = 0$ denotes the lowest-order contribution and $i = 1$ the
\onel\ correction.

The tree-level amplitude involves the $WW$-fusion channel (left diagram
of \reffi{fig:tree}) and the Higgs-strahlung process (right diagram of
\reffi{fig:tree}) where the virtual $Z$~boson is connected to two
electron neutrinos. As explained above, we include the Higgs
propagator corrections into our lowest-order matrix element. We use
\begin{equation}
\cM^{(0)}_{\Phi,e} = \cM^{\rm tree}_{\Phi,e} + \cM^{\rm WF}_{\Phi,e}\,,
\end{equation}
where $\cM^{\rm tree}_{\Phi,e}$ is the contribution of the two tree-level
diagrams, parametrized with $\al = \al_{\rm tree}$, \refeq{alphatree},
and $\cM^{\rm WF}_{\Phi,e}$ denotes the wave-function normalization
contributions given in \refeqs{eq:WWhct2}--(\ref{eq:zhh}) (and
analogously for the $ZZ\{h,H\}$ vertices).

At \onel\ order ($i = 1$), the diagrams shown in
\reffis{fig:WWhHvert}--\ref{fig:enuWCT} contribute (and corresponding
diagrams for the Higgs-strahlung process), involving fermion and sfermion
loops.  The counter-term contributions given in \refeqs{eq:WWhct},
(\ref{eq:WWHct}) enter via the $WW\{h,H\}$ vertices (and analogously for
the $ZZ\{h,H\}$ vertices), while the other counter-term contributions
have the same form as in the SM.

In order to evaluate the cross section that is actually observed in the
detector, $e^+e^- \to (h, H \; + \; {\rm missing~energy})$,
we furthermore take into account the 
amplitude of the 
Higgs-strahlung process where the $Z$~boson is connected to 
$\nu_f \bar \nu_f$ ($f = \mu, \tau$),
\begin{equation}
\label{eq:treeWF}
\cM^{(i)}_{\Phi,f} = \cM^{\rm tree}_{\Phi,f} + 
                     \cM^{\rm WF}_{\Phi,f}, 
\quad (\Phi = h,H;\ i = 0, 1;\ f = \mu,\tau)\,.
\end{equation}
Of course there is no interference between the $\cM^{(i)}_{\Phi,f}$
for different flavors.

For all flavors, on the other hand, the $Z$-boson propagator connected to the
two outgoing neutrinos can become resonant when integrating over the
full phase-space, and therefore a width has to be included in
that propagator. We have incorporated this by using the running width
in the $Z$-boson propagators, $\Ga_Z(s) = (s/\MZ^2) \, \Ga_Z^{\rm exp}$,
where $\Ga_Z^{\rm exp} = 2.4952 \gev$, and dropping the imaginary
parts of the light-fermion contributions to the $Z$-boson
self-energies.  

The cross-section formulas for $h$~production thus become
\begin{align}
\label{sih0}
\si^0_h &\propto \sum_{f=e,\mu,\tau} |\cM^{(0)}_{h,f}|^2 \, , \\
\si^1_h &\propto \sum_{f=e,\mu,\tau} \KL |\cM^{(0)}_{h,f}|^2 
               + 2\, \re \bigl[ (\cM^{(0)}_{h,f})^*\, \cM^{(1)}_{h,f} \bigr]
               \KR \, .
\label{sih1}
\end{align}
The formulas for $H$~production are analogous, except that we have also
included the square of the one-loop amplitude.  This is because the
decoupling behavior of the $WWH$~coupling can make the tree-level cross
section 
very small so that the square of the one-loop amplitude
becomes of comparable size:
\begin{align}
\label{siH0}
\si^0_H &\propto \sum_{f=e,\mu,\tau} |\cM^{(0)}_{H,f}|^2 \, , \\
\si^1_H &\propto \sum_{f=e,\mu,\tau} \KL |\cM^{(0)}_{H,f} \, 
               + \, \cM^{(1)}_{H,f}|^2 \KR \, .
\label{siH1}
\end{align}
In this way, 
at \order{\al^2} only contributions 
$\sim (\cM^{(0)}_{H,f})^*\, \cM^{(2)}_{H,f}$ are neglected, 
which are expected to be very small.


\section{Numerical evaluation}
\label{sec:numeval}

For the numerical evaluation we followed the procedure outlined in
\citere{fa-fc-lt}: The Feynman diagrams for the contributions mentioned
above were 
generated using the \FA~\cite{feynarts} package.  The only necessary
addition was the implementation of 
the counter terms for the $VV\{h,H\}$ ($V = W, Z$) vertices,
\refeqs{eq:WWhct}--(\ref{eq:rcs}), into the existing MSSM model
file~\cite{fa-mssm}.  
The resulting amplitudes were algebraically simplified using \FC\
\cite{formcalc} 
and then automatically converted to a Fortran program.  The \LT\ package
\cite{formcalc, ff} was used to evaluate the one-loop scalar and
tensor integrals. 
The numerical results presented in the following subsections were
obtained with 
this Fortran program.%
\footnote{The code will be made available at {\tt
www.hep-processes.de}.}

While we have obtained results both for the total cross sections and
differential distributions, in the numerical examples below we will
focus on total cross sections only.

To cross-check our tree-level result and the kinematics, we successfully
reproduced the figures of \citere{eennHtree,eennHtree2} and furthermore
performed a detailed comparison with the authors of \cite{logan&su}, who
computed \eeneneH\ using an effective Born approximation. 
We found full agreement within the numerical uncertainties.
We also compared our SM tree-level and \onel\ results with
\citere{didi} and found perfect agreement.
In addition, we checked the phase-space 
integration by comparing the results of three different integration
methods (VEGAS~\cite{vegas}, DCUHRE \cite{dcuhre}, and a stacked
Gaussian integration). 

For our results given below, the following numerical values of 
the SM parameters are used (all other quark and lepton masses are
negligible):
\begin{equation} 
\begin{aligned}
G_F &= 1.16639\times 10^{-5}, &\quad
m_\tau &= 1777 \mev, \\
\MW &= 80.450 \gev, &
\mt &= 174.3 \gev, \\
\MZ &= 91.1875 \gev, &
\mb &= 4.25 \gev, \\
\Gamma_Z &= 2.4952 \gev, &
m_c &= 1.5 \gev .
\end{aligned} 
\end{equation}

In order to fix our notation for the SUSY parameters, we give here
the mass matrix relating 
the $\StopL$ and $\StopR$ states to the mass eigenstates (analogously
for $\Sbot$, $\Stau$, and $\Sneut$)
\begin{equation}
\label{stopmassmatrix}
{\cal M}^2_{\Stop} =
  \ML \msusy^2 + \mt^2 + \CZb (\edz - \frac{2}{3} \sw^2) \MZ^2 &
      \mt \Xt \\
      \mt \Xt &
      \msusy^2 + \mt^2 + \frac{2}{3} \CZb \sw^2 \MZ^2 
  \MR ,
\end{equation}
where the $X_{t,b,\tau}$ read $\Xt = \At - \mu/\tb$,
$X_{b,\tau} = A_{b,\tau} - \mu \, \tb$. Here $A_{t,b,\tau}$
denote the trilinear Higgs--$\Stop,\Sbot,\Stau$ couplings, and 
$\mu$ is the supersymmetric Higgs mass parameter. The mass matrices
for the first two generations of sfermions are defined analogously.
For our numerical
evaluation we have chosen for simplicity a common soft SUSY-breaking 
parameter in the
diagonal entries of the sfermion mass matrices, $\msusy$, and the
same trilinear couplings for all generations. 
Our analytical result, however, holds for general values of the
parameters in the sfermion sector.

The further SUSY parameters entering our result via the Higgs boson
propagator corrections are 
the SU(2) gaugino mass parameter, $M_2$, (the U(1)
gaugino mass parameter is obtained via the GUT relation, 
$M_1 = (5/3)\, (\sw^2/\cw^2)\, M_2$), and the gluino mass, $\mgl$.

For our numerical analyses we assume all 
soft SUSY-breaking parameters to be real.
Our analytical result, however,
holds also for complex parameters entering the loop corrections to
\eennhH.


\subsection{SM Higgs-boson production}
\label{subsec:SMprod}

For comparison purposes, we start our analysis with the fermion-loop
corrections to the process
$e^+e^- \to \bar \nu \nu \HSM$
in the SM.

\begin{figure}[ht!]
\begin{center}
\includegraphics{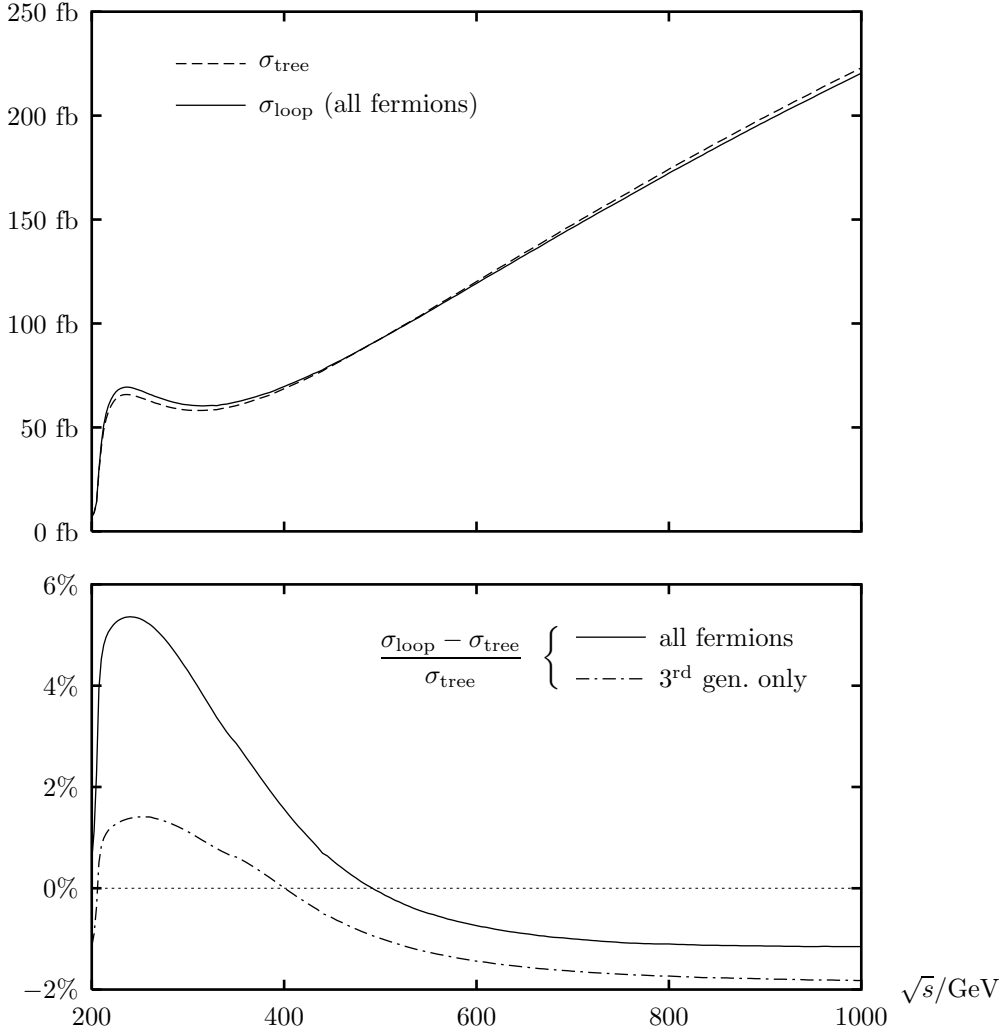}
\caption{
The tree-level and the one-loop cross sections for 
$e^+e^- \to \bar \nu \nu \HSM$ in the SM,
$\si_{\HSM}^0$ and $\si_{\HSM}^1$, are shown as a function of
$\sqrt{s}$ for  $\MHSM = 115 \gev$. The upper plot shows the absolute
values, the lower plot shows the relative corrections for all fermions
and for the third-generation fermions only.
}
\label{fig:HSM_sqrts}
\end{center}
\vspace{-1em}
\end{figure}

\reffi{fig:HSM_sqrts} shows the tree-level and \onel-corrected
production cross section for an SM Higgs-boson mass of 
$\MHSM = 115 \gev$. The absolute values are shown in the upper
plot. The sharp rise in the cross section for $\sqrt{s} \gtrsim 200$~GeV
is due to the threshold for on-shell production of the $Z$~boson in the
Higgs-strahlung contribution, see the right diagram of \reffi{fig:tree}.
Above the threshold the $1/s$ behavior of the Higgs-strahlung
contribution competes with the logarithmically rising $t$-channel
contribution from $WW$ fusion.

The lower plot shows the relative correction coming from all fermions,
as well as the correction from 
the third-generation fermions only. The correction from all
fermions ranges from about
$+5\%$ at low $\sqrt{s}$ to $-1.2\%$ at high $\sqrt{s}$.
Restricting to the contribution of third-generation fermions only,
we obtain corrections in the range
from $+1.3\%$ to $-1.8\%$. 
These corrections (both from the third family only as well as from the 
first two generations) are at the level of the expected sensitivity for the
$WW$-fusion channel at the LC.
For a LC running in its high-energy mode
with $\sqrt{s} \approx 800$~GeV in particular a
measurement of the total cross section with an accuracy of better than
2\% seems to be feasible~\cite{eennHexp}.

\begin{figure}[ht!]
\begin{center}
\includegraphics{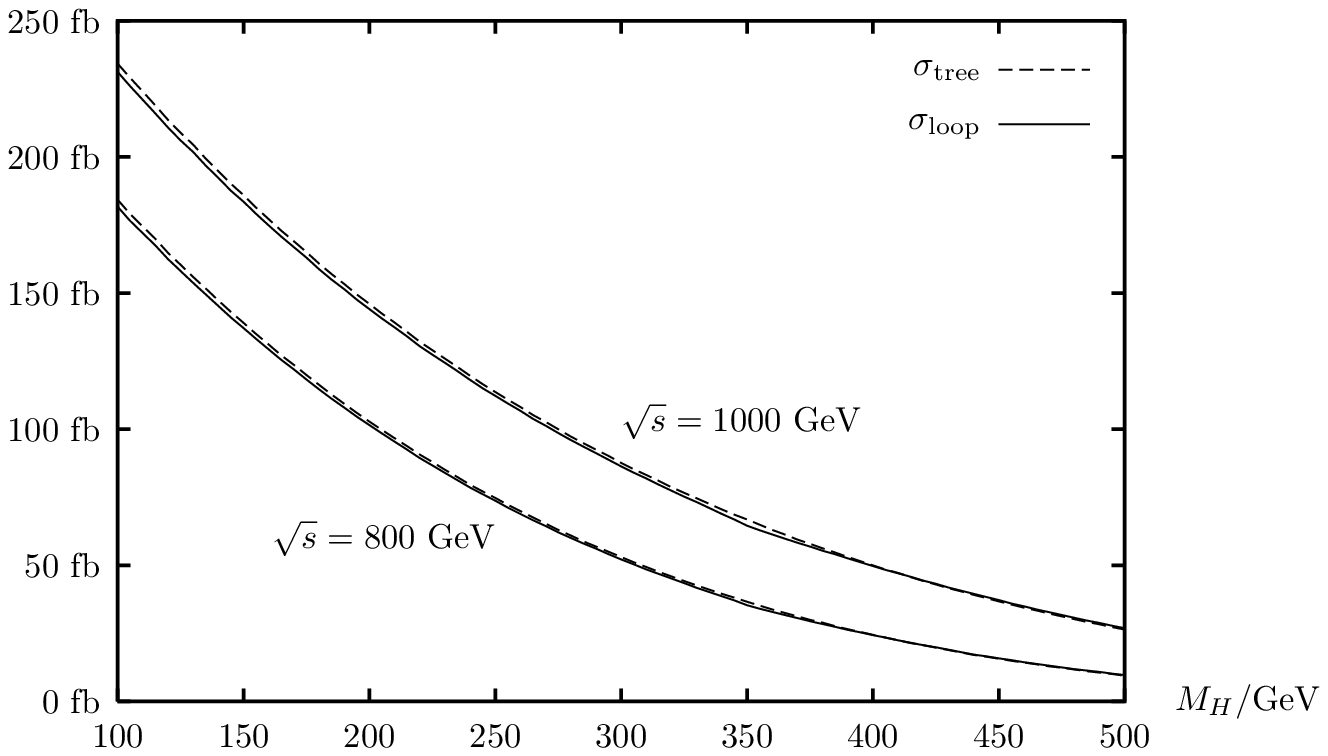}
\caption{
The tree-level and the one-loop cross sections for 
$e^+e^- \to \bar \nu \nu \HSM$ in the SM,
$\si_{\HSM}^0$ and $\si_{\HSM}^1$, are shown as a function of
$\MHSM$ for $\sqrt{s} = 800, 1000 \gev$. 
}
\label{fig:HSM_MH}
\end{center}
\end{figure}

In \reffi{fig:HSM_MH} the SM production cross section is shown
as a function of $\MHSM$ for $\sqrt{s} = 800, 1000 \gev$. An SM Higgs
boson possesses a relatively large production cross section,
\order{10\fb}, depending on the available energy, even for 
$\MHSM \gtrsim 500 \gev$. Thus it should easily be detectable at a 
high-luminosity LC.


\subsection{Light $\cp$-even Higgs-boson production}
\label{subsec:hprod}

Since the mass of the lightest $\cp$-even Higgs boson in the MSSM is
bounded from above by $\Mh \lesssim 135 \gev$~\cite{mhiggslong,mhiggsAEC},
its detection at the LC is guaranteed~\cite{lc}.  In order to exploit
the precision measurements possible at the LC, a precise prediction at
the percent level of its
production cross section (and its decay rates) is necessary.

In the following we analyze the $h$~production cross
section. To begin with, we focus on the four benchmark
scenarios defined in \citere{LHbenchmark} (proposed for MSSM Higgs-boson 
searches at hadron colliders and beyond). $\MA$ and $\tb$ are kept as
free parameters. The four benchmark scenarios are
(more details can be found in \citere{LHbenchmark}) 
\begin{itemize}

\item
the $\mhmax$ scenario, which
yields a maximum value of $\Mh$ for given $\MA$ and $\tb$,
\begin{equation}
\begin{aligned}
{}& \mt = 174.3 \gev, \quad \msusy = 1 \tev, \quad
\mu = 200 \gev, \quad M_2 = 200 \gev, \\
{}& \Xt = 2\, \msusy, \quad
\Atau = \Ab = \At, \quad \mgl = 0.8\,\msusy\,,
\end{aligned}
\label{mhmax}
\end{equation}

\item
the no-mixing scenario, with no mixing in the $\Stop$~sector,
\begin{equation}
\begin{aligned}
{}& \mt = 174.3 \gev, \quad \msusy = 2 \tev, \quad
\mu = 200 \gev, \quad M_2 = 200 \gev, \\
{}& \Xt = 0, \quad
\Atau = \Ab = \At, \quad \mgl = 0.8\,\msusy\,,
\end{aligned}
\label{nomix}
\end{equation}

\item
the ``gluophobic-Higgs'' scenario, with a suppressed $ggh$ coupling,
\begin{equation}
\begin{aligned}
{}& \mt = 174.3 \gev, \quad \msusy = 350 \gev, \quad
\mu = 300 \gev, \quad M_2 = 300 \gev, \\
{}& \Xt = -750 \gev, \quad
\Atau = \Ab = \At, \quad \mgl = 500 \gev\,,
\end{aligned}
\label{gluophobicH}
\end{equation}

\item
the ``small-$\aeff$'' scenario, with possibly reduced decay rates for
$\hbb$ and $\htautau$,
\begin{equation}
\begin{aligned}
{}& \mt = 174.3 \gev, \quad \msusy = 800 \gev, \quad
\mu = 2.5 \, \msusy, \quad M_2 = 500 \gev, \\
{}& \Xt = -1100 \gev, \quad 
\Atau = \Ab = \At, \quad \mgl = 500 \gev\,.
\end{aligned}
\label{smallaeff}
\end{equation}

\end{itemize}
As explained above, for the sake of simplicity, $\msusy$ is chosen as a
common soft SUSY-breaking parameter for all three generations.

\begin{figure}[ht!]
\begin{center}
\vspace{-1em}
\includegraphics[width=17.7cm]{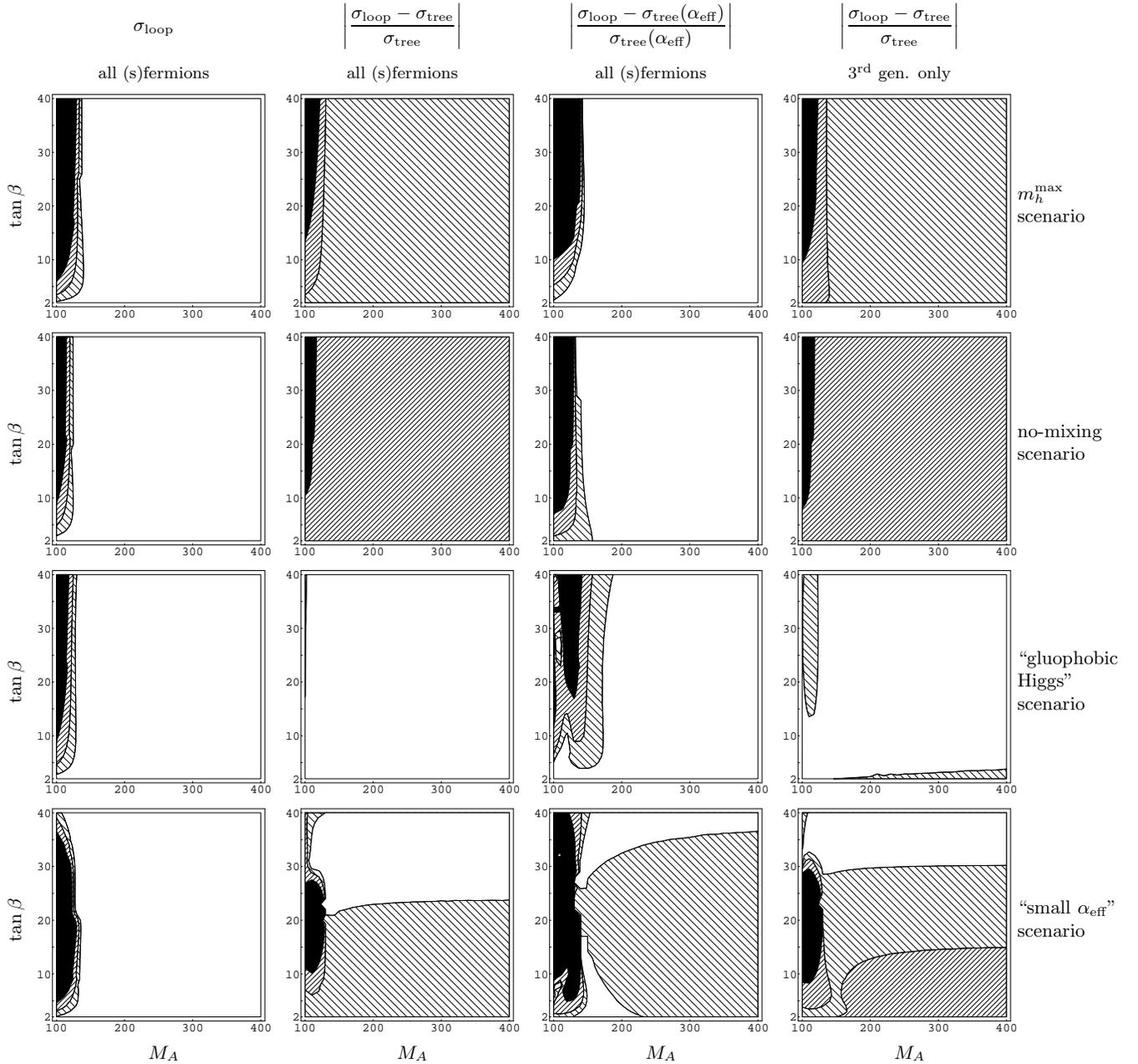}
\caption{
Results for the cross section for \eennh\ at one-loop order,
$\si^1_h$ (left column), the relative corrections including all three
fermion and sfermion generations (two middle columns), and the relative
corrections including only the third generation (right column) are shown
in the $\MA$--$\tb$ plane for four benchmark scenarios at
$\sqrt{s} = 1 \tev$. 
The universal Higgs propagator corrections have
been absorbed into the tree-level cross section and thus do not appear
in the relative corrections. In the second and fourth column 
the Higgs propagator corrections are implemented according 
to \refeqs{eq:treeWF}, (\ref{sih0}), while in
the third column the $\aeff$ approximation is used.
In the results for
$\si^1_h$ (left column) the black region corresponds to 
$\si^1_h < 50 \fb$, the dark-shaded region to
$50 \fb \le \si^1_h \le 100 \fb$, the light-shaded region to
$100 \fb \le \si^1_h \le 150 \fb$, and the white region to 
$\si^1_h \ge 150 \fb$. 
For the relative corrections (second to fourth column) the black region
corresponds to  $R^1_h > 5\%$, the dark-shaded region to
$2\% \le R^1_h \le 5\%$, the light-shaded region to
$1\% \le R^1_h \le 2\%$, and the white region corresponds to 
$R^1_h \le 1\%$ (see text). 
}
\label{fig:h_benchmark}
\vspace{-2em}
\end{center}
\end{figure}

\reffi{fig:h_benchmark} shows in the four benchmark scenarios the
$h$~production cross section, $\si^1_h$, \refeq{sih1}, as well as the
relative size of the loop 
corrections,
\begin{equation}
R^1_h = \Bigg| \frac{\si^1_h - \si^0_h}{\si^0_h} \Bigg|
\label{relcor}
\end{equation}
(which is nearly always negative), 
including the contributions from all generations as well as from the
third generation only. 
The results are shown in the
$\MA$--$\tb$ plane for $100 \gev \le \MA \le 400 \gev$ and 
$2 \le \tb \le 50$. For larger $\MA$ values the behavior of the
lightest $\cp$-even MSSM Higgs boson is very SM-like, i.e.\ the
results hardly vary with $\MA$ any more. 

For very low values of $\MA$, $\MA < 150 \gev$, the cross section is
relatively small. This is due to the fact that the $WWh$~coupling at
tree level, being $\sim \Sba$, can become very small. In this region
of parameter space, however, the heavy $\cp$-even Higgs boson is still 
very light
and couples to the gauge bosons with approximately SM strength, the 
tree-level coupling of
$WWH$ being $\sim \Cba \approx 1$.

For the interpretation of the middle and right column of
\reffi{fig:h_benchmark} it is important to 
keep in mind that we have absorbed the universal Higgs propagator
corrections, which are numerically very important, into our tree-level
cross section. Thus, the relative corrections, $R^1_h$, shown in
\reffi{fig:h_benchmark}, display the effects of the other genuine
one-loop corrections only. We first compare the corrections in the 
two cases where the Higgs propagator corrections are implemented
according to \refeqs{eq:treeWF}, (\ref{sih0}), which ensures the correct
on-shell properties of the outgoing Higgs boson (second column), and
where an $\aeff$ approximation is used (third column), which is often
done in the literature. In the $\aeff$ approximation, the leading
contribution of the process-independent corrections entering via the
Higgs-boson propagators is included by replacing
the tree-level coupling of $\{h,H\}WW = \{\sin,\cos\}(\be - \al)$
by $\{\sin,\cos\}(\be - \aeff)$.
The difference between the full on-shell prescription
and the $\aeff$ approximation turns out to be sizable. It amounts to
several percent even for relatively large values of $\MA$. As a
consequence, including the Higgs propagator corrections in an $\aeff$
approximation will not be sufficient in view of prospective precision
measurements of the \eennh\ cross section.

The results in the second column of \reffi{fig:h_benchmark} show that
the size of the corrections from fermion and sfermion loops is somewhat
different in the four scenarios. While corrections of more than 5\%
only occur for $\MA \lesssim 130 \gev$, we obtain corrections of 2--5\% in
the whole parameter space of the no-mixing scenario. 
Corrections of 1--2\% can be found in large  
parts of the parameter space of the $\mhmax$ and the small-$\aeff$ scenario.
The situation in the four benchmark scenarios, which have been chosen to
represent different aspects of MSSM phenomenology, shows that the
corrections investigated here are typically of the order of about 1--5\%.
A measurement of the \eennh\ cross section at the percent level will
thus be sensitive to this kind of corrections. 

In the right column of \reffi{fig:h_benchmark} we show $R^1_h$ derived
including the contributions from the third family of fermions and
sfermions only. Thus the differences between the second and the fourth
column reflect the relevance of the loop corrections coming from the
first two families. While in the $\mhmax$~and the no-mixing scenario
differences can mostly be found for small $\MA$, $\MA \lesssim 150 \gev$,
in the other two scenarios the effect of the first two families can be
relevant also for larger $\MA$. 
Within the gluophobic-Higgs scenario, the first two families play a
role for small $\tb$ and large $\MA$. In the small-$\aeff$ scenario
differences can be found for larger $\tb$ over the whole $\MA$ range.
In the latter two scenarios, the corrections coming from the first and
second family lead to a partial compensation of the corrections from the
third family. The light fermion generations can give rise to a
contribution of $\sim 1\%$, which is non-negligible for
cross section measurements at the percent level.

The Higgs propagator corrections, which we have absorbed into our 
tree-level cross section, mainly affect the numerical value of $\Mh$,
which enters the final-state kinematics, while the numerical effect of
the corrections to the 
$WWh$ coupling is less important. The comparison between the
prediction for \eennh\ in the MSSM and the corresponding process in the
SM for the same value of the Higgs boson mass (which is not shown here)
yields deviations of more than 5\% for $\MA \lesssim 200 \gev$, which to a
large extent are due to the suppressed $WWh$~coupling in the MSSM case. 
Deviations of more than 1\% are found in all scenarios up to rather 
large values of $\MA$.

\begin{figure}[ht!]
\begin{center}
\includegraphics{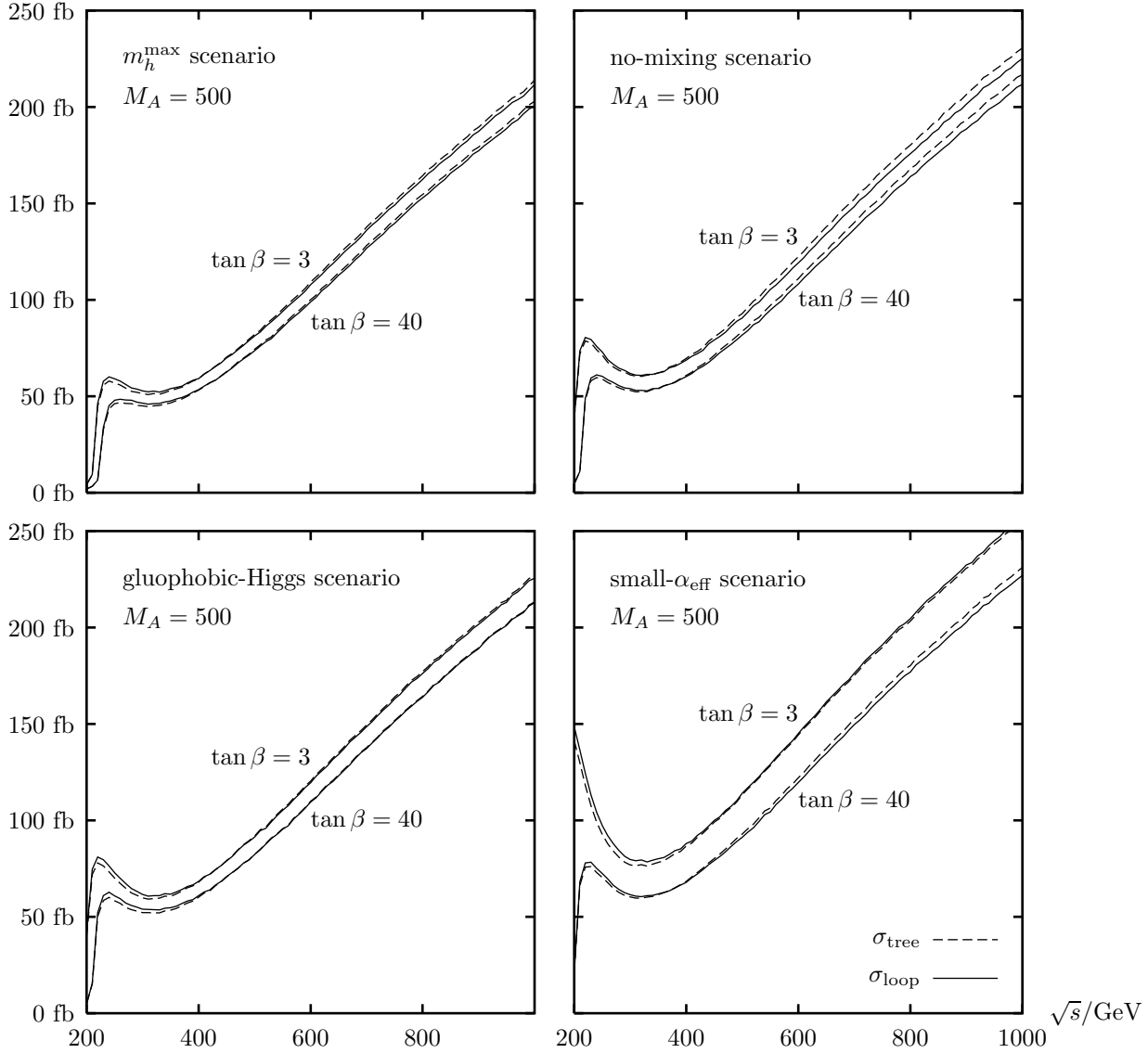}
\caption{
The tree-level and the one-loop cross sections for 
\eennh,
$\si_h^0$ and $\si_h^1$, are shown as a function of $\sqrt{s}$ in the
four benchmark scenarios for $\MA = 500 \gev$ and
$\tb = 3, 40$. 
}
\label{fig:sih_sqrts}
\end{center}
\end{figure}

\reffi{fig:sih_sqrts} shows our results for \eennh\ in the four 
benchmark scenarios as a function of $\sqrt{s}$ for 
$\MA = 500 \gev$ and $\tb = 3, 40$. Note that the
difference in the cross sections for the four benchmark scenarios for
given $\MA$ and $\tb$ is entirely due to SUSY loop corrections
(which, as explained above, affect in particular the value of $\Mh$).

\begin{figure}[ht!]
\begin{center}
\includegraphics{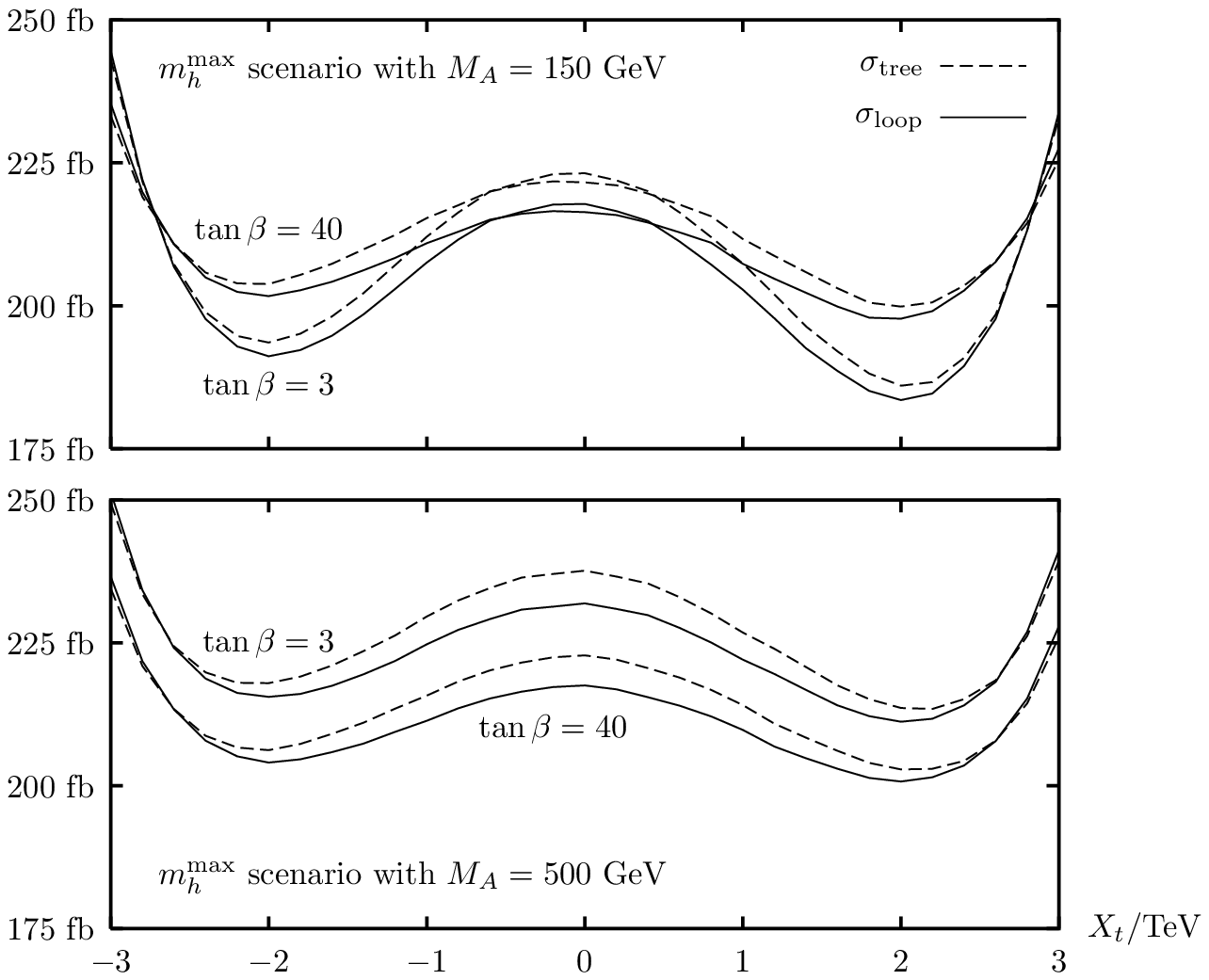}
\caption{
The tree-level and the one-loop cross sections for 
\eennh,
$\si_h^0$ and $\si_h^1$, are shown as a function of $\Xt$.
In the upper (lower) plot $\MA$ has been set to $\MA = 150$ $(500) \gev$, 
$\tb$ is fixed at $\tb = 3, 40$. The other parameters are as given 
in \refeq{mhmax}.
}
\label{fig:sih_Xt}
\end{center}
\end{figure}

The numerically important effects of the Higgs propagator corrections
become apparent in particular 
from \reffi{fig:sih_Xt}, where the tree-level 
and the one-loop
cross sections are shown as a function of $\Xt$, i.e.\ the mixing in the
scalar top sector. The plots are given for the four combinations of 
$\MA = 150, 500 \gev$ and $\tb = 3, 40$, and the other parameters 
(besides $\Xt$) 
are chosen as in the 
$\mhmax$ scenario. The variation of the tree-level cross sections
indicates the effect of the Higgs propagator corrections affecting both
the value of $\Mh$ and the 
Higgs coupling to gauge bosons. These
corrections can change the cross section by up to $\sim 25\%$,
while the other loop corrections typically stay below 2.5\%.

\begin{figure}[ht!]
\vspace{3em}
\begin{center}
\includegraphics{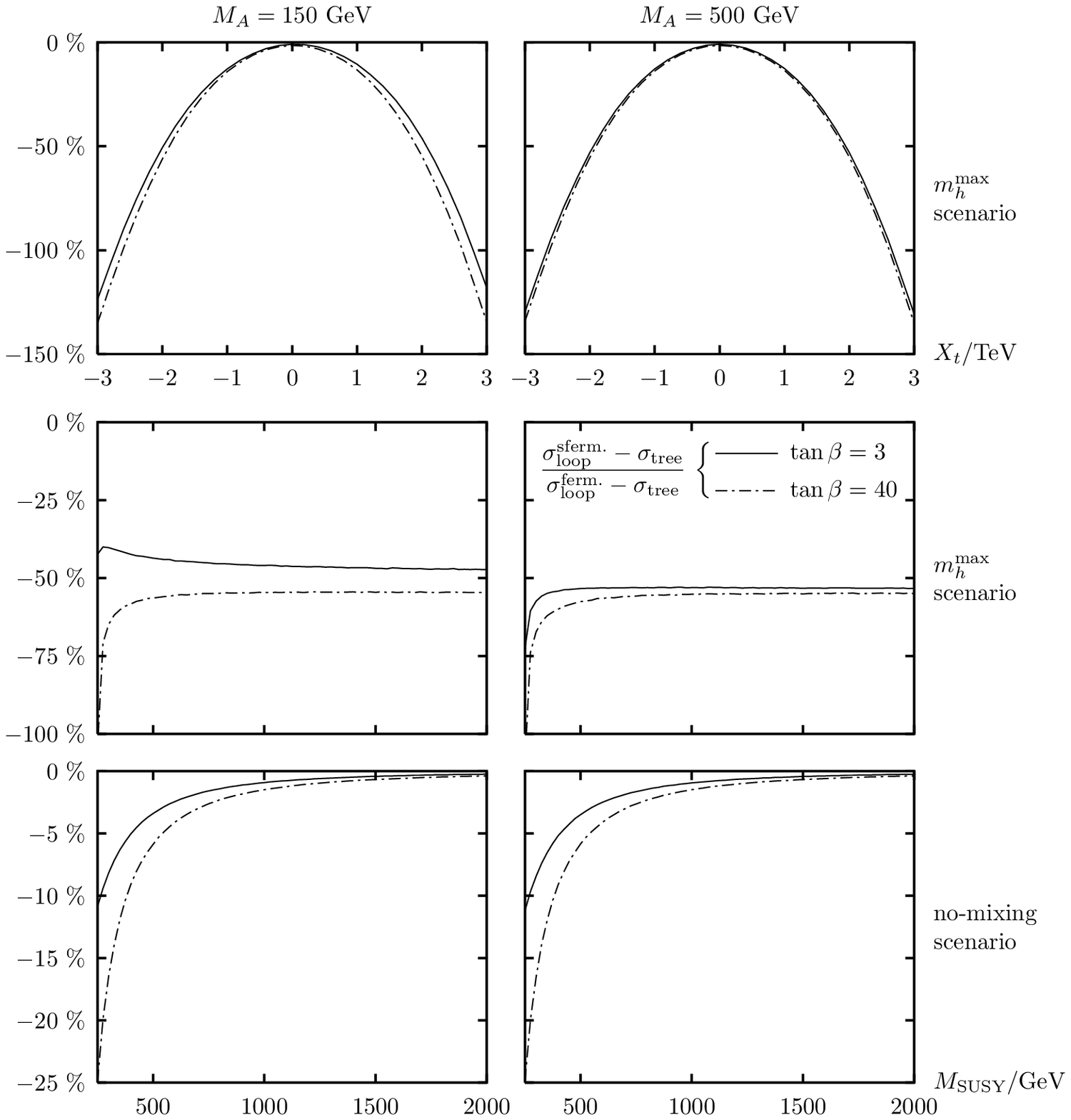}
\vspace{1em}
\caption{
The relative size of the purely sfermionic loop corrections to $\si^1_h$
as compared to the purely fermionic \onel\ corrections is shown as a 
function of $\Xt$ (upper) and $\msusy$ (middle and lower row) for all
combinations  
of $\MA = 150, 500 \gev$ and $\tb = 3, 40$. The other parameters are
chosen as in the $\mhmax$~(upper and middle) or as in the no-mixing
scenario (lower row). 
}
\label{fig:sih_sf_effects}
\end{center}
\vspace{3em}
\end{figure}

Finally, in \reffi{fig:sih_sf_effects} we analyze the relative
importance of the purely sfermionic loop corrections (corresponding to
the Feynman diagrams with sfermion loops in 
\reffis{fig:WWhHvert}--\ref{fig:enuWCT}; as before, the Higgs propagator
corrections absorbed into the tree-level result contain both fermion- and
sfermion-loop contributions). These corrections
constitute, as explained earlier, a UV-finite and gauge-invariant
subset of the loop contributions. The relative size of the sfermion
corrections as compared to the purely fermionic \onel\ corrections 
is shown in \reffi{fig:sih_sf_effects}. The upper row shows the
relative size as a function of $\Xt$ in the 
$\mhmax$ scenario for all combinations of $\MA = 150, 500 \gev$ and
$\tb = 3, 40$. While for $\Xt \approx 0$, i.e.\ for small splitting in
the scalar top sector, the sfermionic corrections are small, their
contribution becomes more important for increasing $|\Xt|$.
They have the opposite sign of
the purely fermionic corrections and thus partially compensate
their effects. For
$|\Xt/\msusy| \approx 2$ the sfermionic corrections are about 
half as large
as the fermion corrections. For very large $|\Xt|$ (which also lowers
$\Mh$ substantially) they can become even bigger than the fermionic ones.

In the lower part of \reffi{fig:sih_sf_effects} (middle and lower row)
we analyze the relative size of the purely sfermionic corrections in
the $\mhmax$~(middle) and the no-mixing scenario (lower row) as a
function of $\msusy$.  
In the no-mixing scenario, for increasing $\msusy$ the relative size
of the sfermion corrections 
becomes smaller, as can be expected in the decoupling
limit~\cite{decoupling1l,decoupling2l}.
In the $\mhmax$ scenario, however, the situation is different. Here
$\Xt \approx \At$ is fixed to $\At \approx \Xt = 2 \, \msusy$. In the
$h\Stope\Stopz$ vertex, being $\sim \At\Ca + \mu\Sa$, the coupling is
proportional to the SUSY mass scale. This results in a term 
$\sim \At/\msusy$ in the \onel\ corrected $WWh$ vertex, which for
large $\msusy$ goes to a constant and can be of the order of the
purely fermionic correction. The cross section then behaves as 
$\sim \Xt^2/\msusy^2$ as can be seen in the upper row of
\reffi{fig:sih_sf_effects}. 

\subsection{Heavy $\cp$-even Higgs-boson production}
\label{subsec:Hprod}

We now investigate the effects of loop corrections on the cross section
for heavy $\cp$-even Higgs-boson production in the MSSM. As explained
above, these corrections are of particular interest in the decoupling
region, i.e.\ for large values of $\MA$.
If $\MA \lesssim \sqrt{s}/2$, the heavy Higgs bosons can be pair-produced at
the LC via $e^+e^- \to Z^* \to HA$. Beyond this kinematical limit,
$H$~production is in principle possible via the $WW$-fusion and the
Higgs-strahlung channels. This production mechanism is heavily
suppressed at tree-level, however, owing to the decoupling property of
the $H$~coupling to gauge bosons. If loop-induced contributions turn out
to be sizable in the mass range $\MH > \sqrt{s}/2$, an enhanced reach of
the LC for $H$~production could result.

\begin{figure}[ht!]
\vspace{5em}
\mbox{} \hspace{5em}
\includegraphics{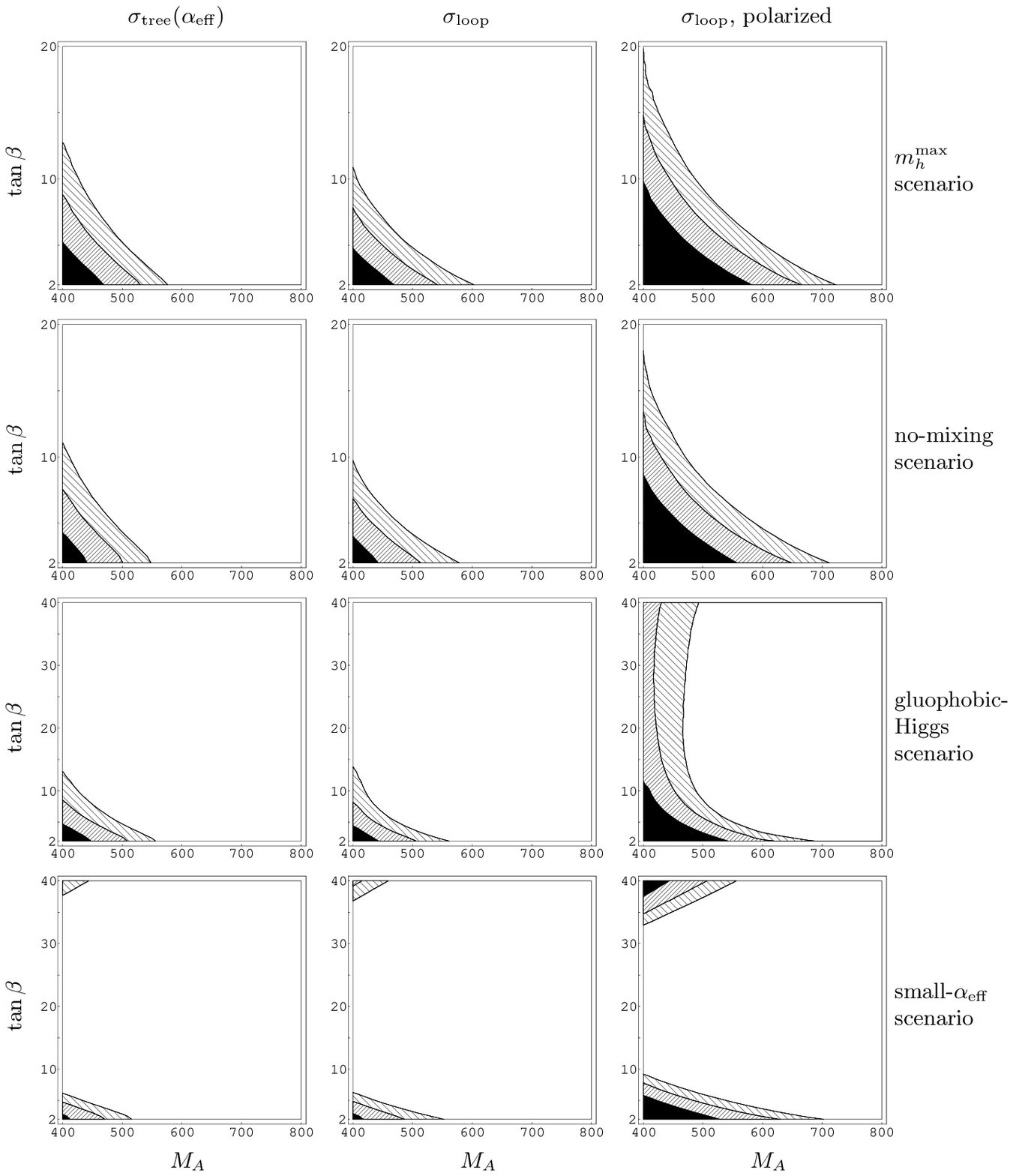}
\vspace{1em}
\caption{
The tree-level cross section for \eennH, $\si^0_H(\aeff)$, 
using an $\aeff$~approximation (left), and the one-loop cross 
sections, $\si^1_H$, without (middle), and with beam polarization
(right column), are shown in the $\MA$--$\tb$ plane for four benchmark
scenarios for $\sqrt{s} = 1 \tev$. The different shadings correspond to:
white: $\si \le 0.01 \fb$, light shaded: $0.01 \fb \le \si \le 0.02 \fb$, 
dark shaded: $0.02 \fb \le \si \le 0.05 \fb$, black: $\si \ge 0.05 \fb$.
}
\label{fig:H_benchmark}
\end{figure}

In \reffi{fig:H_benchmark} we first compare the tree-level cross section
evaluated in the $\aeff$~approximation, $\si_H^0(\aeff)$ (left column),
and the one-loop cross section according to \refeq{siH1}, $\si_H^1$
(middle column), in the four benchmark scenarios. 
For phenomenological analyses of MSSM Higgs-boson production in this 
channel the cross
section has so far mostly been evaluated using an $\aeff$~Born
approximation (see also \refse{subsec:hprod}).

We concentrate on the case of $\sqrt{s} = 1 \tev$. Since we are
interested in $\MH \approx \MA > \sqrt{s}/2 = 500 \gev$, we focus on
the region $400 \gev \le \MA \le 900 \gev$, and scan over the whole $\tb$
region. For a LC like TESLA, the anticipated integrated luminosity is
of \order{2 \iabm}. For this luminosity, 
a production cross section of about $\si_H = 0.01 \fb$ constitutes a
lower limit for the observation of the heavy $\cp$-even Higgs boson.
(For the scenarios discussed below, the dominant decay channel of $H$
is the decay into $t \bar t$ or $b \bar b$, depending on the value of
$\tb$, and also sizable branching ratios into SUSY particles are 
possible in some regions of parameter space; a more detailed simulation of 
this process should of course take into account the impact of the decay
characteristics on the lower limit of observability, while in this work
we use the approximation of a universal limit.)
This area is shown in white in \reffi{fig:H_benchmark}. 
In the four benchmark scenarios shown in \reffi{fig:H_benchmark}, the
inclusion of the loop corrections that go beyond the $\aeff$~Born
approximation turns out to have only a moderate effect on the
area in the $\MA$--$\tb$ plane in which $H$~production could be
observable. While for the $\mhmax$ and the no-mixing scenario the area
is slightly decreased to smaller $\tb$ values and somewhat enlarged to
higher $\MA$ values, the area is slightly decreased in $\tb$ in the
gluophobic-Higgs scenario (and stays approximately the same in $\MA$),
while the area is slightly enlarged both in $\MA$ and $\tb$ in 
the small-$\aeff$ scenario. 
For the four benchmark scenarios, an
observation with $\MH > 500 \gev$ is only possible for low $\tb$, 
$\tb \lesssim 5$, where the LC reach in $\MH$ can be extended by up to 
$100 \gev$. It should be noted at this point that while the area of
observability is modified only slightly in the plots as a consequence of 
including the loop corrections, the relative changes between the
tree-level and the one-loop values of the cross sections can be very large,
owing to the suppressed $WWH$ coupling in the tree-level cross section.

The prospects for observing a heavy Higgs boson beyond the kinematical
limit of the $HA$ pair production channel become more favorable, 
however, if polarized beams are
used. The cross section becomes enhanced for left-handedly polarized
electrons and right-handedly polarized positrons. While a 100\%
polarization results in a cross section that is enhanced roughly by a
factor of 4, more realistic values of 80\% polarization for
electrons and 60\% polarization for positrons~\cite{polarization}
would yield roughly an enhancement by a factor of 3. 
The right column of \reffi{fig:H_benchmark} shows the four benchmark
scenarios with 100\% polarization of both beams. The area in the 
$\MA$--$\tb$ plane in which observation of the $H$~boson might become
possible is strongly increased in this case. In the
$\mhmax$ and the no-mixing scenario, $H$ observation could be possible
for small $\tb$ up to $\MA \lesssim 700 \gev$. In the gluophobic-Higgs
and the small-$\aeff$ scenario this effect is somewhat smaller. In the
latter scenario the discovery of a heavy Higgs boson in the parameter
region $\MA > 500 \gev$ will become possible also for large $\tb$, $\tb
\gtrsim 35$.

While without the inclusion of beam polarization we do not find a
significant enhancement of the LC discovery reach as a consequence of 
the loop corrections in the four benchmark scenarios analyzed in
\reffi{fig:H_benchmark}, this behavior changes in other
regions of the MSSM parameter space. As a particular example, we
investigate the $\MA$--$\tb$ plane in the ``$\siHenh$'' (``enhanced cross
section'') scenario, which is defined by
\begin{equation}
\msusy = 350 \gev, \quad \mu = 1000 \gev, 
\label{xsmax}
\end{equation}
with the other parameters as in the $\mhmax$ scenario, \refeq{mhmax}.
The $\siHenh$ scenario is characterized by a relatively small value of $\msusy$
and a relatively large value of~$\mu$. Large $\tb$ values, $\tb \gtrsim 30$, 
can result in low and experimentally ruled-out $\Sbot$~masses and are
therefore omitted.

\begin{figure}[ht!]
\begin{center}
\includegraphics{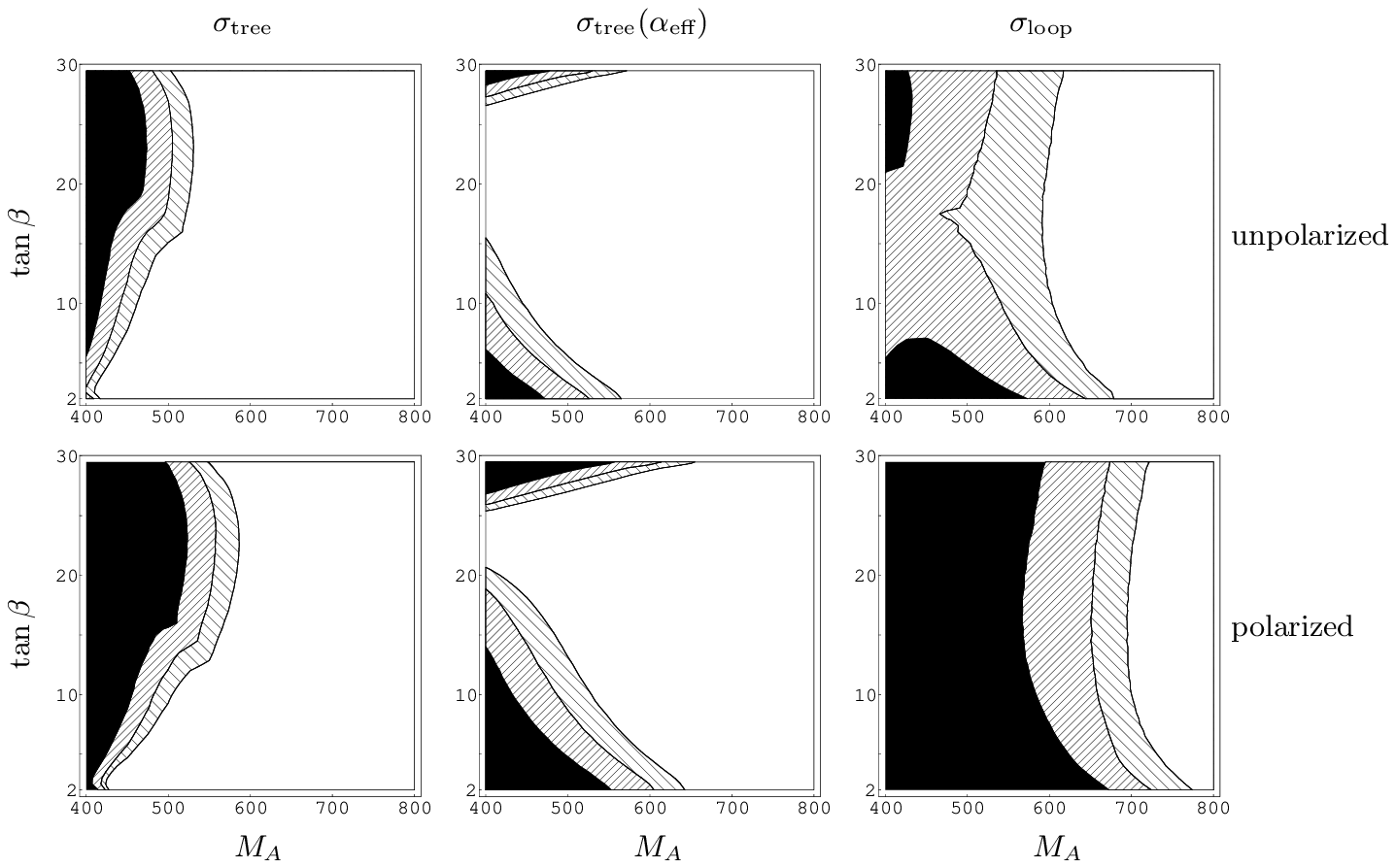}
\caption{
The cross sections for \eennH\
are shown in the $\MA$--$\tb$ plane for the $\siHenh$ scenario, 
\refeq{xsmax}. The tree-level cross section (left) including the
finite wave-function corrections is compared to the
$\aeff$~approximation (middle) and the \onel\ corrected cross section
(right column). 
The upper (lower) row shows the production cross section for
unpolarized (100\% polarized) electron and positron beams. 
The color coding is as in \reffi{fig:H_benchmark}.
}
\label{fig:xsmax}
\end{center}
\end{figure}

\begin{figure}[ht!]
\begin{center}
\includegraphics{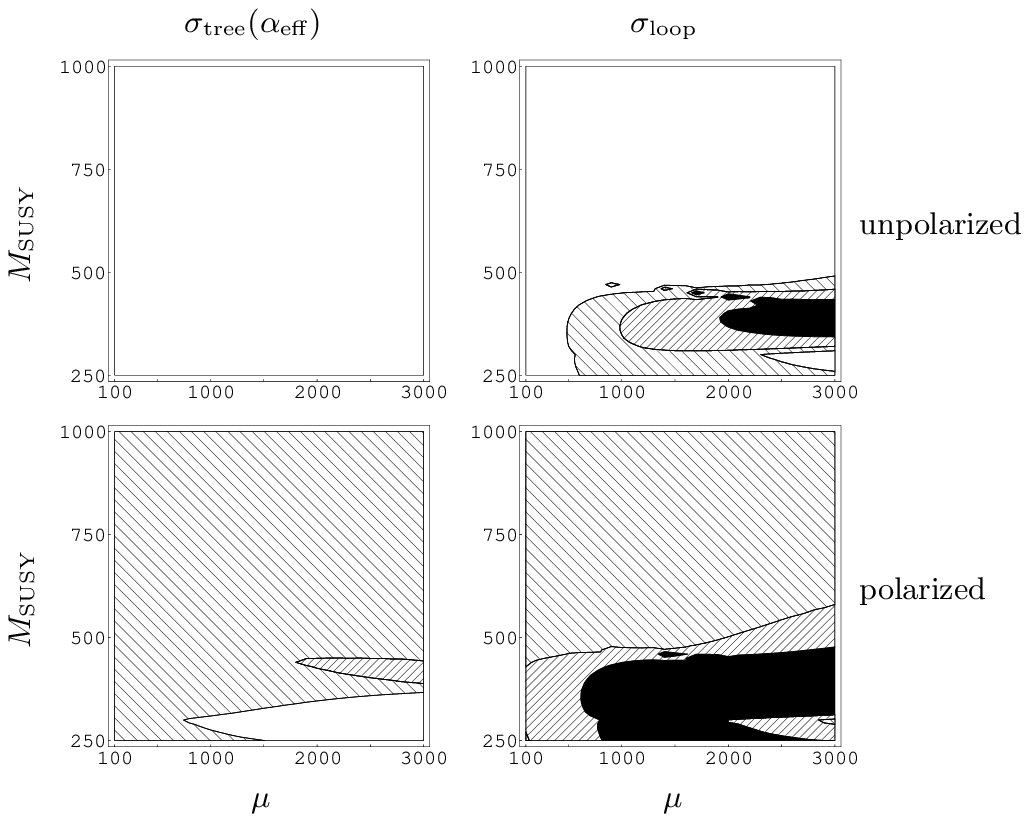}
\caption{
The tree-level cross section for \eennH, $\si^0_H(\aeff)$, 
using an $\aeff$~approximation (left), and the one-loop cross 
section, $\si^1_H$ (right column), is 
shown in the $\mu$--$\msusy$ plane for the parameters of
\refeq{mhmax} and $\MA = 600 \gev$ and $\tb = 4$. The upper row shows
the unpolarized case, while in the lower row the effects of beam
polarization are included. The color coding
is as in \reffi{fig:H_benchmark}.
}
\label{fig:mumsusy}
\end{center}
\end{figure}

In the upper row of \reffi{fig:xsmax} we compare $\si_H^0$,
$\si_H^0(\aeff)$, and $\si_H^1$ for the parameters of the
$\siHenh$ scenario, \refeq{xsmax}, in the unpolarized case. 
The figure shows that both the inclusion of the finite Higgs propagator
wave-function corrections as compared to the $\aeff$ approximation 
(left vs.\ middle column) and of the genuine one-loop corrections
(right vs.\ left column) is very important in this scenario. 
According to the $\aeff$ Born approximation, the parameter area in which 
observation of $H$ is possible would not be significantly larger than in the
benchmark scenarios discussed in \reffi{fig:H_benchmark}. For $\MA \gtrsim
500$~GeV, observation of the $H$ boson is only possible for either rather
small, $\tb \lesssim 5$, or rather large, $\tb \gtrsim 28$, values of $\tb$.
Inclusion of the finite Higgs propagator wave-function corrections,
which ensure the correct on-shell properties of the outgoing Higgs
boson, changes the situation considerably. While for small and large
values of $\tb$ the additional corrections suppress the \eennH\ cross section,
restricting the $H$ observability to values of $\MA$ below 500~GeV,
observation of the heavy $\cp$-even Higgs boson of the MSSM becomes possible for 
$\MA\lesssim 550$~GeV for a significant range of intermediate values of $\tb$. 
Including also the genuine one-loop
corrections (right column) leads to a further drastic enhancement of the
parameter space in which the $H$ boson could be observed. The
observation of the $H$ boson will be possible in this scenario for all
values of $\tb$ if $\MA \lesssim 600$~GeV, i.e.\ the discovery reach of
the LC in this case is enhanced by about 100~GeV compared to the 
$HA$ pair production channel.

The prospects in this scenario for observation of the heavy $\cp$-even 
Higgs boson of the MSSM become even more favorable if
polarized beams are used.
The lower row of \reffi{fig:xsmax} shows the situation with 100\%
polarization of both beams for the 
$\siHenh$ scenario, \refeq{xsmax}. While in this case the
tree-level result (both for the case including the finite wave-function 
corrections and for the $\aeff$~approximation) gives rise to
observable rates for $\MA \lesssim 600 \gev$ for a certain range of $\tb$
values only, the further genuine loop
corrections enhance the cross section significantly. In this situation
the observation of the heavy $\cp$-even Higgs boson might be possible for
values of $\MA$ up to about 700--750~GeV for all $\tb$ values,
corresponding to an enhancement of the LC reach by more than $200 \gev$.
Cross-section values in excess of $0.05 \fb$ are obtained in this example
for all $\tb$ values for $\MA \lesssim 600 \gev$.

In order to investigate whether this result is a consequence of a very
special choice of SUSY parameters or a more general feature, in
\reffi{fig:mumsusy} we choose the parameters of the $\mhmax$ scenario for
a fixed combination of $\MA$ and $\tb$, $\MA = 600 \gev$, $\tb = 4$, but
scan over $\mu$ and $\msusy$.  The choice $\MH \approx \MA = 600 \gev$
implies that the $HA$ production channel is clearly beyond the reach of a
1-TeV LC. The upper row of \reffi{fig:mumsusy} shows that according
to the tree-level cross section (using the $\aeff$ approximation) an
observable rate for a heavy $\cp$-even Higgs boson with $\MA = 600 \gev$
cannot be found for any of the scanned values of $\mu$ and $\msusy$.
Inclusion of the further loop corrections changes this
situation significantly and gives rise to observable rates
in this example for nearly all $\msusy \lesssim 500 \gev$ if 
$\mu \gtrsim 500 \gev$. 
The visible ``structure'' at $\msusy \approx 500 \gev$ is the result
of several competing effects that affect the finite Higgs
wave-function corrections. 

The same analysis, but with 100\% polarization of both beams, 
is shown in the lower
row of \reffi{fig:mumsusy}. The tree-level $\aeff$~approximation
results in an observable rate in nearly the whole plane apart
from the area with $\mu \gtrsim 1000 \gev$ and $\msusy \lesssim 350 \gev$.
Adding the loop corrections again improves the
situation. No unobservable holes remain in the $\mu$--$\msusy$ plane,
i.e.\ the heavy $\cp$-even Higgs boson with $\MH \approx 600 \gev$
should be visible at a 1-TeV LC with (idealized) beam polarization
in this scenario. The
production cross section is larger than 
$0.02 \fb$ for all $\mu$ and $\msusy \lesssim 500 \gev$ and mostly even
larger than $0.05 \fb$ for $\mu \gtrsim 500 \gev$.

\begin{figure}[ht!]
\vspace{-1em}
\mbox{} \hspace{3em}
\includegraphics{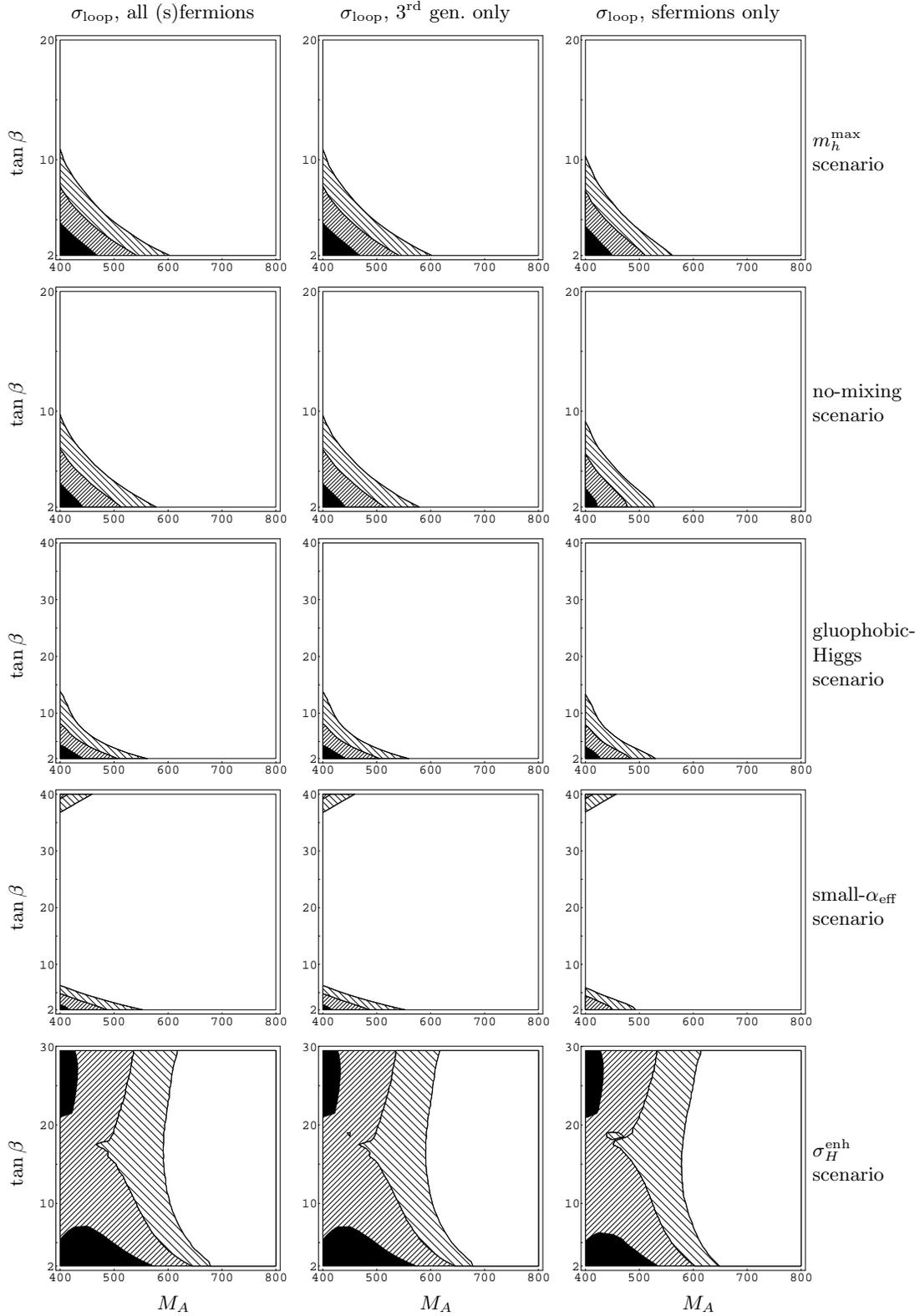}
\caption{
$\si_H^1$ is shown in the $\MA$--$\tb$ plane for the four benchmark scenarios
defined in \refeqs{mhmax}--(\ref{smallaeff})
and the $\siHenh$ scenario for $\sqrt{s} = 1 \tev$ in the unpolarized
case. The \onel\ result containing the corrections from all fermion and
sfermion loops (left) is compared with the result from only
third-generation fermions and sfermions (middle) and the purely
sfermionic corrections from all families (right column). 
The color coding is as in \reffi{fig:H_benchmark}.
}
\label{fig:H_benchmark_contrib}
\vspace{-1em}
\end{figure}

In \reffi{fig:H_benchmark_contrib} we compare for all five scenarios
analyzed in this paper the results for the cross section when different
parts of the generic one-loop corrections are taken into account. In all
cases the full result for the Higgs propagator corrections being
absorbed in the lowest-order cross section are employed.
The left column shows the result containing the
corrections from all fermion and sfermion loops, $\si_H^1$, which is
repeated from previous plots for comparison purposes.
The middle column shows the cross-section prediction based on 
taking into account only the corrections
from the third-generation fermions and sfermions,
which have turned out to be the leading
corrections for the light Higgs-boson production. The result shown in the
right column have been obtained by including only the purely sfermionic
contributions from all generations, i.e.\ the fermion-loop corrections
are omitted in this case. As expected from \reffi{fig:H_benchmark},
in the four scenarios defined in \refeqs{mhmax}--(\ref{smallaeff}),
where the loop corrections turned out to modify the parameter regions
in which $H$ observation becomes possible only slightly, omitting the 
contributions of the first two generations of fermions and sfermions 
and of the fermion loops
of all three generations does not lead to a qualitative change in the
$H$ discovery reach. In the $\siHenh$ scenario, on the other hand, the 
genuine one-loop corrections had a considerable impact on the area in
the $\MA$--$\tb$ plane in which $H$ observation is possible, see 
\reffi{fig:xsmax}. The result displayed in the middle column for this
scenario shows that the bulk of the corrections comes from the third
generation of fermions and sfermions, i.e.\ omission of the first two
generations does not lead to significant effects in the $\MA$--$\tb$ plane.
The result in the right column for this scenario shows furthermore that
the omission of all fermion-loop corrections leads only to very moderate
changes of the parameter regions where $H$ observation is possible. As a
consequence, the by far dominant corrections in this scenario can be
identified as the ones from the sfermions of the third generation.
This is contrary to the $h$~production, where we found that the
fermionic corrections are mostly larger than the sfermionic ones.


\subsection{Comparison with existing results}
\label{subsec:comp}

We have calculated the corrections to the production of both $\cp$-even
Higgs bosons of the MSSM, $h$ and $H$, via the process \eennhH.
The results for the $h$~production can be compared with the existing
result from \citere{Wiener}. 

Our calculation differs from the one in \citere{Wiener} in several respects. 
Our renormalization differs from the one used in \citere{Wiener} for
the Higgs-boson field-renormalization constants and for $\tb$. 
Concerning the field-renormalization constants, we incorporate a finite
wave-function correction that ensures the correct on-shell properties of
the outgoing Higgs boson, while in \citere{Wiener} the Higgs propagator
corrections have been taken into account in an $\aeff$ approximation
only.
Concerning $\tb$, we use the \msbar\ definition that in general leads
to a better numerical stability~\cite{MSSMren,doink}.
While in \citere{Wiener} the contributions from the Higgs-strahlung
diagrams have been taken into account in lowest order only, we have
incorporated also the loop corrections for this class of diagrams in
the same way as for the $WW$-fusion diagrams. 
Furthermore we have evaluated the corrections from the fermion and 
sfermion loops of all generations, while in \citere{Wiener} only the
third generation has been taken into account. In \refse{subsec:hprod}
it has been shown that relevant corrections can also arise from 
the first two generations.

We have performed a numerical comparison with the results given in
\citeres{Wiener,Eberl} for the SM and light $\cp$-even Higgs
production in the MSSM, restricting our results to the contributions
of the third generation only.
\begin{itemize}

\item

Our tree-level results show a rather pronounced threshold
peak for small values of $\sqrt{s}$, where on-shell production of both
the Higgs and the $Z$~boson becomes possible. In this region our results
largely differ from those of \citere{Wiener}, while we find very good
agreement with the results of \citere{didi}. For large values of
$\sqrt{s}$, i.e.\ far above the threshold, our tree-level results
roughly agree with the ones of \citere{Wiener}.

\item
For the loop corrections in the SM case we find significant deviations
with the results of \citere{Wiener} over the whole parameter space.
While the authors of \citere{Wiener} find large corrections of up to
$-15\%$, the corrections in our result for the SM cross section 
turn out to be moderate and do not exceed $\pm 2\%$ (see
\reffis{fig:HSM_sqrts}--\ref{fig:HSM_MH}). The size of the corrections
that we find agrees quantitatively with the estimate of the leading
term in the  
heavy-top expansion~\cite{knisp}. As a further cross-check of our
results, we compared the SM result at the tree and at the \onel\ level 
(for the third generation only and for all generations) with the one 
of \citere{didi} and found perfect agreement.

\item
The same qualitative difference compared to the results of
\citere{Wiener} that we find for the fermion-loop
corrections in the SM also occurs for the $h$ production cross section
in the MSSM. Again, the fermion-loop corrections in our result turn out
to be much smaller than the ones in \citere{Wiener}. Related to this
fact, the relative importance of the corrections from fermion and
sfermion loops is different in our result. While in \citere{Wiener} the
relative size of the purely sfermionic
corrections does not exceed $+11\%$ of the full correction, we find that
the sfermionic corrections can become as large as the purely fermionic
ones for large mixing in the scalar top sector.

\item
In addition to the above-mentioned large deviations, which cannot arise
from the use of slightly different renormalization schemes and of
different approximations, there are further differences related to the
fact that we have implemented the Higgs propagator corrections according
to \refeqs{eq:treeWF}, (\ref{sih0}), while in \citere{Wiener} an $\aeff$
approximation has been used. As shown in \reffi{fig:h_benchmark}, 
the $\aeff$ approximation differs from the full on-shell result by several
percent and is therefore not sufficiently accurate in view of the
prospective precision reachable at a LC in this channel.

\end{itemize}


\section{Conclusions}
\label{sec:conclusinos}

We have investigated the production of the $\cp$-even MSSM Higgs-bosons
at a future Linear Collider in the process \eennhH. This process is mediated 
via the $WW$-fusion mechanism, which dominates at higher energies, and via
the Higgs-strahlung process, which is important at low energies. We have
evaluated all one-loop contributions from fermions and sfermions, and we
have furthermore implemented the numerically large process-independent
Higgs-boson propagator corrections in such a way that the correct
on-shell properties of the outgoing Higgs bosons are ensured.

At a high-energy Linear Collider, the process \eennh\ will be the production 
mode with the highest cross section. For the genuine loop corrections 
beyond those arising from Higgs propagator contributions, we have found 
corrections of about 1--5\%. These corrections will be relevant in view
of the anticipated precision of the cross-section measurement at the LC.
The same is true for the deviations between the result containing the 
full Higgs-boson propagator corrections and the result based on the
$\aeff$ approximation.

We have also evaluated the correction for the corresponding SM
Higgs-production process and found for $\MHSM = 115 \gev$ corrections
in the range of $+5\%$ for small $\sqrt{s}$, 
$\sqrt{s} \approx 250 \gev$, to $-1.2\%$ for large $\sqrt{s}$.

Restricting our result to the contributions from third-generation
fermions and sfermions only (and disregarding the correction arising
from replacing the $\aeff$ approximation by the full Higgs-boson
propagator corrections in the MSSM case), we have compared our results
for the SM case and for light $\cp$-even Higgs-boson production in the
MSSM with the corresponding results given in \citere{Wiener}. We find
significant deviations both in the overall size of the radiative
corrections, which we find to be considerably smaller, and in the
relative importance of fermion- and sfermion-loop contributions.

Concerning the production of the heavy $\cp$-even Higgs boson, we find
that in a set of benchmark scenarios proposed for Higgs-boson searches at
future colliders, the genuine loop corrections beyond those arising from 
Higgs propagator contributions only slightly enhance the discovery reach 
of the Linear Collider. In the case of polarized electron and positron
beams, however, the Linear Collider reach can be significantly extended
beyond the kinematical limit of the $HA$ pair-production channel. 
In more favorable regions of the MSSM parameter space, the genuine 
loop corrections can drastically enlarge the parameter space for which
detection of the heavy $\cp$-even Higgs boson becomes possible. In such
a scenario, assuming polarized beams, at $\sqrt{s} = 1 \tev$ the
detection of $H$~could be possible up to $\MH \approx 750 \gev$.


\subsection*{Acknowledgements}

We thank W.~Hollik for helpful discussions.
We thank S.~Dittmaier, H.~Logan, and S.~Su
for providing us with detailed numbers of their calculations to
cross-check our results and for useful discussions. 
We furthermore thank H.~Eberl for helpful
communication concerning the comparison of our results. 
G.W.\ thanks the Max-Planck Institut f\"ur Physik in Munich for
the hospitality offered to him during his stay where parts of this work
were carried out.
This work has been supported by the European Community's Human
Potential Programme under contract HPRN-CT-2000-00149 Physics at
Colliders. 



\end{document}